\documentclass[a4paper,11pt]{article}
\pdfoutput=1
\usepackage{jcappub}
\usepackage[T1]{fontenc}
\usepackage[utf8]{inputenc}
\usepackage{array}
\usepackage{paracol}
\usepackage{mathrsfs}
\usepackage{bm}
\usepackage{url}
\usepackage[normalem]{ulem}
\usepackage{mathtools}
\usepackage{array}
\usepackage{orcidlink}
\newcolumntype{P}[1]{>{\centering\arraybackslash}p{#1}}
\newcolumntype{M}[1]{>{\centering\arraybackslash}m{#1}}
\usepackage[caption=false]{subfig}
\usepackage{dcolumn}
\usepackage{graphicx, epsfig}
\usepackage[dvipsnames]{xcolor}
\usepackage{yhmath}
\usepackage[caption=false]{subfig}
\usepackage[normalem]{ulem}
\usepackage{bigints}
\usepackage{float}
\usepackage{hyperref}
\usepackage{booktabs}
\usepackage{longtable}
\usepackage{caption}
\usepackage{supertabular}

\def\t13{\mathrel{{\theta_{13}}}}
\def\y12{\mathrel{{\tan^2 \theta_{12}}}}
\def\c2{\mathrel{{\chi^2 }}}

\newcommand{\be}{\begin{equation}}
\newcommand{\ee}{\end{equation}}
\newcommand{\ba}{\begin{eqnarray}}
\newcommand{\ea}{\end{eqnarray}}

\makeatletter
\renewcommand{\subheader}[1]{%
  \gdef\@subheader{\normalsize\normalfont #1}%
}
\makeatother
\subheader{FERMILAB-PUB-26-0015-T}
\title{
Constraining high-energy neutrinos from tidal disruption events with IceCube high-energy starting events
}
\author[a,b,c,1]{Mainak Mukhopadhyay\orcidlink{0000-0002-2109-5315},\note{Corresponding author.}}
\author[c]{Patrick  Wusinich\orcidlink{0009-0001-2197-3819},}
\author[c,d]{and Kohta Murase\orcidlink{0000-0002-5358-5642}}
\affiliation[a]{Astrophysics Theory Department, Theory Division, Fermi National Accelerator Laboratory, Batavia, Illinois 60510, USA.}
\affiliation[b]{Kavli Institute for Cosmological Physics, University of Chicago, Chicago, Illinois 60637, USA.}
\affiliation[c]{Department of Physics; Department of Astronomy Astrophysics; Center for Multimessenger Astrophysics, Institute for Gravitation and the cosmos, The Pennsylvania State University, University Park, Pennsylvania 16802, USA.}
\affiliation[d]{Center for Gravitational Physics and Quantum Information, Yukawa Institute for Theoretical Physics, Kyoto, Kyoto 606-8502 Japan.}
\emailAdd{mainak@fnal.gov}
\emailAdd{pjw5497@psu.edu}
\emailAdd{murase@psu.edu}
\abstract{
Tidal disruption events (TDEs) have been proposed as candidate sources of high-energy neutrinos. Successful and choked jets, as well as the accretion disk, corona, wind, and outflow regions in a TDE have been examined and shown to produce TeV - PeV neutrinos. In this work, we use the IceCube 12.5 year high energy starting events (HESE) dataset and perform a maximum likelihood analysis to investigate the spatial and temporal correlations between HESE dataset and a selected sample of 89 TDEs. Our results indicate that the currently observed data do not show any significant correlation and hence is consistent with the background only hypothesis. Using this result, we place constraints on the fraction of TDEs harboring intrinsic jets ($f_{\rm jet}$) and the corresponding isotropic-equivalent cosmic ray (CR) energy ($\mathcal{E}_{\rm CR}$). We note that even with limited statistics, we can constrain the parameter space as $\mathcal{E}_{\rm CR} \lesssim 3 \times 10^{53}$ erg for $f_{\rm jet} \gtrsim 0.6$ at more than 90\% C.L. Finally, we discuss the theoretical implications of our results and the limits on the all-sky diffuse neutrino flux from TDEs. With more observational data in the electromagnetic band for TDEs and neutrino observations from IceCube and KM3NeT, our analysis can be used to place stringent constraints on physical parameters associated with TDEs.
}
\begin{document}
\maketitle
\flushbottom
\section{Introduction}
\label{sec:intro}
Tidal disruption events (TDEs) are extreme cosmic phenomena where a star (of mass and radius $M_*$ and $R_*$ respectively) is shred apart by a supermassive black hole (SMBH) (of mass $M_{\rm BH}$) when it approaches sufficiently close to the SMBH, or approximately the tidal disruption radius $R_T \sim R_* (M_{\rm BH}/M_*)^{1/3}$~\cite{Rees1988,Komossa:2015qya,Stone:2018nbx}. Roughly half the debris from the star is lost in an unbound orbit, while the other half eventually falls back and circularizes at a distance $R_{\rm circ}\approx 2 R_T$~\cite{1989ApJ...346L..13E,1989IAUS..136..543P}. A part of the circularized debris may form an accretion disk~\cite{LacyTownesHollenbach1982,Hayasaki:2012ia,Hayasaki:2015pxa,Bonnerot:2016krr,Bonnerot:2019yjb,Bonnerot:2020pyz}. The fallback rate of the debris scales with time as $\dot{M}_{\rm fb} \propto t^{-5/3}$~\cite{Rees1988,1989IAUS..136..543P}. However, at initial times the accretion rate can be similar to the super-Eddington rate. This enhanced accretion rate can result in a magnetized and geometrically thick accretion flow, leading to a pile up of magnetic flux near the SMBH. This pile up of the magnetic flux may lead to the formation of a magnetically-arrested disk (MAD), which can help with launching a jet~\cite{Tchekhovskoy:2013tt}. Furthermore, TDEs accompanied by MADs are likely to produce jets through the Blandford-Znajek mechanism~\cite{Blandford:1977ds,Tchekhovskoy:2009ba,Tchekhovskoy:2011zx,Kelley:2014tga,Liska:2018btr}.

Jetted TDEs\footnote{We define jetted TDEs as jet-loud TDEs constituting powerful successful jets. These are among TDEs that harbor an intrinsic jet, which may or not be successful and observable.} have distinct observational features~\cite{Burrows:2011dn, Zauderer:2011mf,DeColle:2019wzp}. The interaction of the jet with the interstellar medium~\cite{Giannios:2011it} can result in them being radio loud (luminosity $\gtrsim 10^{44}$ erg/s). Additionally, such jets also have X-ray emission with rapid variability. Multiple jetted TDEs, like Swift J1644+57~\cite{Bloom:2011xk,Burrows:2011dn}, Swift J2058+05~\cite{Cenko:2011ys}, Swift J1112-8238~\cite{Brown:2015amy}, and AT2022cmc~\cite{Andreoni:2022afu}, have been observed in various electromagnetic (EM) bands. Observationally, however roughly 1\% of all TDEs are thought to be jetted. Furthermore, the interaction of the jet with the surrounding debris is highly uncertain~\cite{Tchekhovskoy:2013tt}, since the jet launch time and its orientation with the orbital plane is non-trivial. Recent radio observations~\cite{Horesh:2021gvp,Cendes:2023rrc,Sato:2024zpq,Wu:2025vsr} clearly show evidence of a delayed peak in the radio light curve when compared to the optical peak. This hints towards a delayed central engine activity including delayed launching of the jet~\cite{Liska:2018btr,Mukhopadhyay:2023mld}. Thus, jetted TDEs remain a prime target for transient searches and are crucial for gaining insight into non-trivial dynamics of TDEs.

Although TDEs have been observed through photons across the EM spectrum (see~\cite{Gezari:2021bmb} for a review), other messengers from TDEs such as gravitational waves~\cite{Pfister:2021ton,Toscani:2025uar}, neutrinos (see~\cite{Hayasaki:2021jem} for a review), and cosmic rays~\cite{Farrar:2008ex,Farrar:2014yla,Zhang:2017hom,Guepin:2017abw,Piran:2023svv,Plotko:2024gop} are yet to be confirmed. In particular, TDEs are considered to be high-energy neutrino sources~\cite{Berezinsky1977} thanks to multiple sites of particle acceleration. The relativistic jet~\cite{murase2008astrophysical,Dai:2016gtz,Senno:2016bso,Guepin:2017abw,Lunardini:2016xwi,Liu:2020isi,Winter:2020ptf,Murase:2020lnu,Zheng:2022afy,Lu:2023miv}, the accretion disk (in the MAD or RIAF state)~\cite{Hayasaki:2019kjy, Murase:2020lnu}, the disk corona~\cite{Murase:2020lnu}, and the wind or outflow~\cite{Murase:2020lnu,Winter:2022fpf} regions have been identified as sites of high energy neutrino production in the literature. 
A relativistic jet in a TDE can accelerate CR protons and nuclei through shocks and magnetic dissipation to energies between $10^{15} - 10^{18}$ eV. The accelerated CR protons can interact with the soft X-ray photons ($\sim 0.1 - 10$ keV) from the disk or the wind/outflow, or at the internal jet dissipation region through inverse-Compton scattering of soft photons to soft X-rays. In the case of choked jets, the resulting photon field from the trapped radiation enhances the high-energy neutrino production rates. For choked jets, $p\gamma$ interactions are typically dominant for high energy neutrino production.

The relevant acceleration and cooling mechanisms and hence the expected neutrino spectra will vary depending on the region of high energy neutrino production in the TDE. However, a more generic statement can be made about the expected model independent neutrino fluence from a TDE. The fluence will be proportional to the isotropic-equivalent cosmic ray energy ($\mathcal{E}_{\rm CR}$). Assuming CRs accelerated due to shocks or magnetic reconnection have a $E_{\rm CR}^{-2}$ spectra, the expected neutrino fluence from all TDEs can be written as~\cite{Esmaili:2018wnv}
\be
\label{eq:nusig}
N_\nu^{\rm exp} = \sum_{i=1}^{N_{\rm cat}} \sum_{\alpha} \int_{10\ \rm TeV}^{10\ \rm PeV} \frac{1}{8} \frac{1}{4\pi d_{L,i}^2} \frac{\mathcal{E}_{\rm CR}}{\mathcal{R}E_\nu^2} \mathcal{A}_{\rm eff}^{\nu_\alpha} (E_\nu)\ dE_\nu \,,
\ee
where $N_{\rm cat}$ is the total number of TDEs. We sum over the contribution from each of the $i$-th TDE. The second summation accounts for the neutrino flavors $\alpha = \nu_e,\nu_\mu, \nu_\tau$. The factor $1/8$ accounts for the parent proton energy emitted in neutrinos. The luminosity distance of the $i$-th TDE is given by $d_{L,i}$ and the total isotropic-equivalent energy going into cosmic rays is given by $\mathcal{E}_{\rm CR}$. The parameter $\mathcal{R} = {\rm ln}\left( E_{\rm CR}^{\rm max}/E_{\rm CR}^{\rm min} \right)$. We choose $E_{\rm CR}^{\rm max} = 10^9$ GeV and $E_{\rm CR}^{\rm min} = 10$ GeV which gives $\mathcal{R} \sim 18$. We assume a standard $E_\nu^{-2}$ for the neutrino spectra. The flavor-dependent effective area is given by $\mathcal{A}_{\rm eff}^{\nu_\alpha}$ and we use it from~\cite{IceCube:2020wum}. The neutrino energies are integrated between 10 TeV to 10 PeV. In reality, the limits on $\mathcal{E}_{\rm CR}$ will depend on the specific spectral index, but choosing the integration limits between 10 TeV to 10 PeV, makes our assumption well-suited~\cite{Senno:2017vtd}.

High-energy neutrino emission and its association with TDEs have been examined with great interest since there was some evidence of at least three neutrino events coincident with a TDE. These were AT2019dsg~\cite{Stein:2020xhk}, AT2019fdr~\cite{Reusch:2021ztx}, and AT2019aalc~\cite{vanVelzen:2021zsm}\footnote{Additional associations have also been claimed~\cite{Yuan:2024foi}.}. However, recent updates with improved directional reconstruction from IceCube show that the reconstructed neutrino directions differ from the direction of the associated TDEs~\cite{IceCube:2025uzh}. Nevertheless, the possibility of TDEs contributing to the flux of high-energy neutrinos and their eventual detection remain exciting. Apart from single source associations, stacking searches have been performed with Type Ib/c~\cite{Esmaili:2018wnv}, choked jet supernovae~\cite{Senno:2017vtd,Chang:2022hqj}, galaxy mergers~\cite{Bouri:2024ctc}, supermassive black hole binaries~\cite{PugazhendhiAD:2025cjj}, active galactic nuclei (AGNs) along with Seyfert galaxies~\cite{Murase:2019vdl,Kheirandish:2021wkm, IceCube:2024dou, Abbasi:2025tas} to look for associations of high-energy neutrino events using various datasets from IceCube.

The true rate density of TDEs observationally inferred through various EM surveys is given by $\dot{\rho}^{\rm true}_{\rm TDE} \sim 10^{-5} - 10^{-4}\ {\rm gal}^{-1}{\rm yr}^{-1} = 10^2 - 10^3\ {\rm Gpc}^{-3}{\rm yr}^{-1}$~\cite{Magorrian:1999vm,vanVelzen:2014dna,Sun:2015bda,Stone:2020vdg}. The apparent rate density of on-axis jetted TDEs from (mostly X-ray) observations can be estimated to be $\dot{\rho}_{\rm TDE}^{\rm on-axis} \sim 10^{-9} - 10^{-8}\ {\rm gal}^{-1}{\rm yr}^{-1}= 0.01 - 0.1\ {\rm Gpc}^{-3}{\rm yr}^{-1}$~\cite{Bloom:2011xk,Burrows:2011dn,Metzger:2011qq}. Thus, the fraction of TDEs that have jets pointing towards us is $f_{\rm jet}\sim {10}^{-4}-{10}^{-3}$.

In this work, for the first time, we perform a stacking search to correlate the observed HESE dataset from IceCube~\cite{hesedata} with a catalog of observed TDEs~\cite{Langis:2025btl}. The stacking search is performed using an unbinned maximum likelihood function exploring associations in the spatial and temporal domains between the two datasets. Based on the discussions above, we choose two physical parameters of importance -- the fraction of TDEs harboring intrinsic jets ($f_{\rm jet}$) and the total isotropic equivalent energy in CRs ($\mathcal{E}_{\rm CR}$) which are crucial for neutrino production. Even with limited statistics, we obtain limits in the $f_{\rm jet} - \mathcal{E}_{\rm CR}$ plane. These limits can be further translated to constrain the contribution of TDEs to the diffuse high-energy neutrino flux. The present analysis is timely since with the advent of the Vera C. Rubin Observatory Legacy Survey of Space and Time (LSST)~\cite{LSST:2008ijt,Blum:2022dxi}, the number of TDEs detected\footnote{In the era of LSST roughly $10^5$ transients are expected to be seen per day. However, it is important to note that the classification of a transient as a TDE requires revisiting the sky locations in an interval of a few days. Therefore, identifying a transient event as a TDE is highly challenging and in the baseline survey strategy of LSST only a small fraction of the transients can be identified as TDEs~\cite{Bricman:2023jzx}.} are estimated to be $\sim 1000$ per year. Furthermore, with IceCube's continued operation along with the Cubic-Kilometre Neutrino Telescope (KM3NeT)~\cite{KM3Net:2016zxf} and upcoming neutrino observatories like IceCube-Gen2~\cite{IceCube-Gen2:2020qha}, Giant Radio Array for Neutrino Detection (GRAND)~\cite{GRAND:2018iaj}, and Radio Neutrino Observatory in Greenland (RNO-G)~\cite{RNO-G:2020rmc}, the available data from neutrino observatories will also significantly increase. With increased statistics, more stringent constraints can possibly be placed on various physical quantities of interest associated with TDEs.\\

The paper is organized as follows. We discuss the TDE and neutrino catalogs that we use in Sec.~\ref{sec:detcat}. In Sec.~\ref{sec:unbinnedL}, we elaborate on the likelihood analysis pipeline that is used to compute the test statistic. Using the results from the previous section, we place upper limits on physical properties of the model, the fraction of TDEs with jets ($f_{\rm jet}$), and the energy budget in cosmic rays ($\mathcal{E}_{\rm cr}$) in Sec.~\ref{sec:lim}. We conclude and discuss the implications of our work in Sec.~\ref{sec:concl}.
\section{Data catalogs}
\label{sec:detcat}
In this section, we outline the details associated with our TDE catalog and the IceCube HESE neutrino dataset. With the currently available data, we are indeed limited by low statistics. This problem, however, will be remedied easily with observational data for TDEs from the Vera C. Rubin Observatory (LSST) and additional HESE data release from the IceCube Collaboration.
\subsection{TDE catalog}
\label{subsec:tde_cat}
\begin{figure}
    \centering
    \includegraphics[width=1\linewidth]{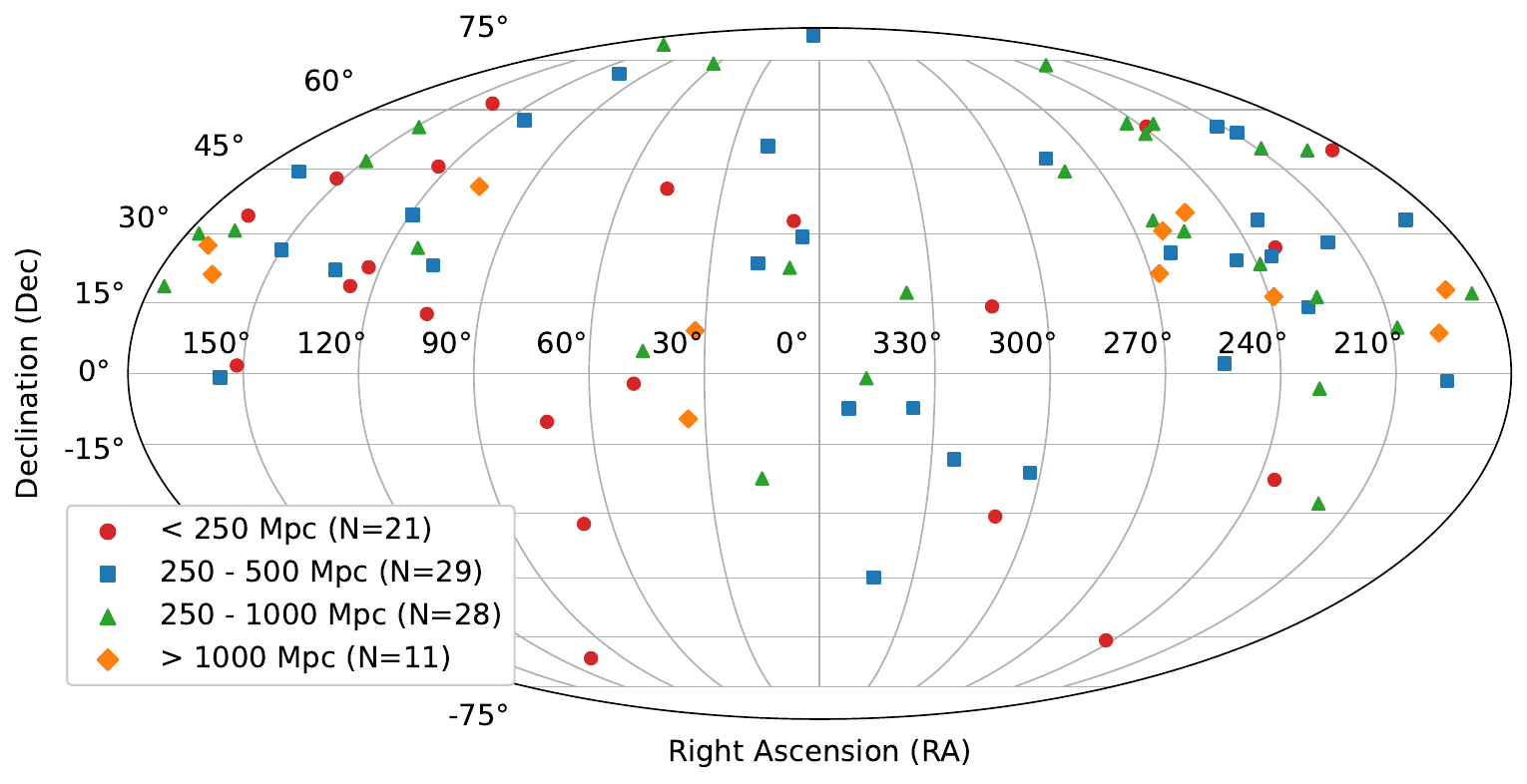}
    \caption{\label{fig:tdes}Mollweide projection for the TDEs in the catalog~\cite{Langis:2025btl} that we select for our analysis (see the text for details). The colors and markers denote the luminosity distance of the TDEs.}
\end{figure}
The TDE sample used for this analysis comes from a catalog presented in~\cite{Langis:2025btl}, which includes publicly available photometry data sourced from the Black Hole Target Observation Manager (BHTOM)\footnote{BHTOM (Black Hole Target and Observation Manager)  \url{https://bh-tom2.astrouw.edu.pl/about/}}. This catalog provides the right ascension (RA) and declination (Dec) coordinates, redshift ($z$), and discovery date for 134 TDEs. The data sourced from BHTOM provides time-series photometry including the observation time and magnitude in a specific band with uncertainty. Besides these quantities, we also need the time of peak brightness $t_{\rm peak}$. Below we discuss the algorithm to compute $t_{\rm peak}$ from the provided data. The relevant details about the telescope and the corresponding bands used and the computed value of $t_{\rm peak}$ is discussed in Appendix~\ref{appsec:tpk_tde_catalog} and Table~\ref{tab:tde_band_grouping}.

\begin{figure}
\centering
\includegraphics[width=0.49\linewidth]{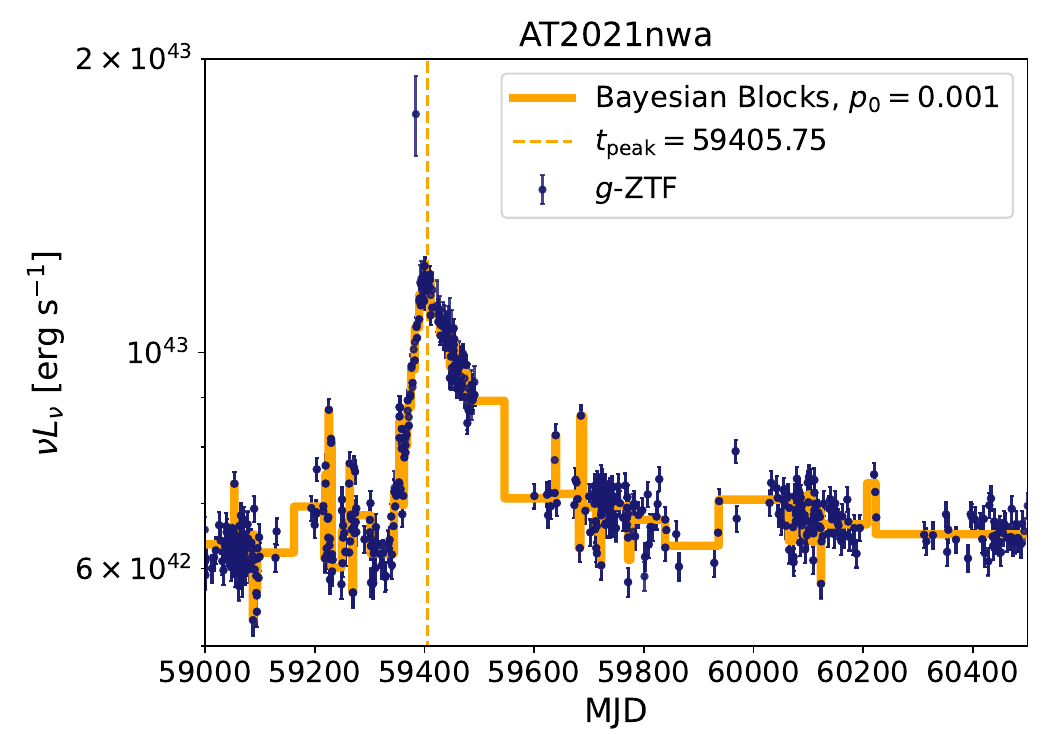}
\includegraphics[width=0.49\linewidth]{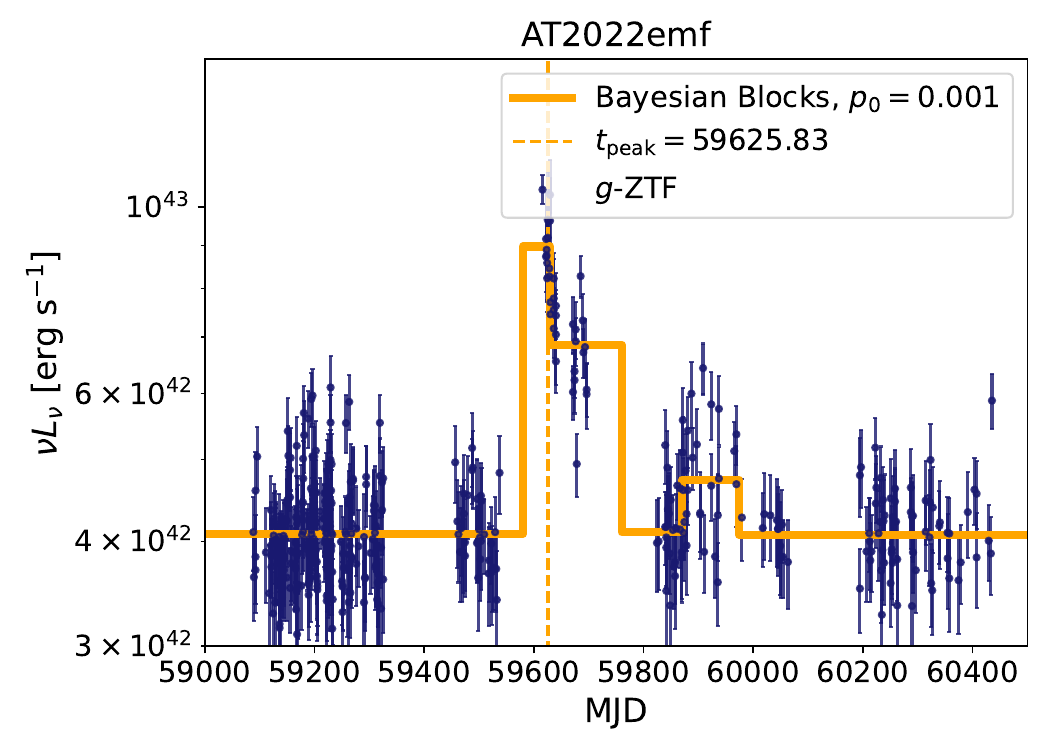}
\caption{\label{fig:bb_tde_example}Examples of the Bayesian Blocks segmentation algorithm applied to the ZTF luminosity light curves of two randomly selected TDEs from the catalog. The dark blue points show the ZTF photometry with error bars, while the solid orange curves show the optimal Bayesian-block representation of the light curves. The dashed orange vertical line marks the midpoint of the data points contained within the peak block, which we use to define $t_{\rm peak}$.
}
\end{figure}

Our technique of computing $t_{\rm peak}$ from the given photometric data is the same as presented in Ref.~\cite{Langis:2025btl}. The band magnitudes from the photometry data are converted to a normalized relative flux using the monotonic transform $f_i \propto 10^{-0.4\,mag_i}$ with uncertainties propagated as $\sigma_{f,i} = 0.4\,\ln(10)\, f_i\, \sigma_{i}$ when photometric uncertainties are available. Otherwise, we estimate a Gaussian uncertainty from the flux scatter. Following Ref.~\cite{Langis:2025btl}, we implement a Bayesian Blocks segmentation of this flux time series with a globally optimal set of change points. Through this implementation, we compute the mean relative flux for each block, which corresponds to the maximum-likelihood estimate under a Gaussian log-likelihood fit. This Bayesian Blocks change-point segmentation accounts for model complexity through a penalty on additional change points and prevents overfitting \cite{Scargle:1998, Scargle:2013}. The block with the highest mean relative flux is defined as the \textit{peak block}. We set the time of peak brightness to the midpoint of the earliest and latest measurement times within this peak block. Alternatively, we found the MJD of the brightest data point under the same band-selection procedure outlined above, but requiring only one measurement within the time window. The time of peak brightness that we obtain from using the Bayesian Blocks algorithm is broadly consistent with those reported by Ref.~\cite{Hammerstein:2022wia} (smaller sample of 30 TDEs) for the cases where the brightest episode contained within a single Bayesian Block. However for cases, where the peak block is only partially populated by observational data points, indicating that the true peak lies outside the observed baseline, the resulting midpoint estimate can be systematically offset from the true peak time. Under this framework, approximations for the time of peak brightness are more adaptive to irregular cadence and noisy outliers than simply choosing the single brightest data point. Therefore, we set the time of peak brightness ($t_{\rm peak}$) using this Bayesian Blocks algorithm in this analysis.

In Fig.~\ref{fig:bb_tde_example}, we show two randomly selected TDEs as examples of the Bayesian-block procedure used to estimate $t_{\rm peak}$. In this procedure, the parameter $p_0$ sets the nominal false-alarm probability used to determine the prior penalty for adding additional change points; smaller values of $p_0$ therefore produce fewer, broader blocks and reduce the chance of over-segmenting noise fluctuations. We adopt $p_0=10^{-3}$ throughout this analysis (and checked that our results are convergent for smaller $p_0$ values than the one adopted). The ZTF photometry is converted to $\nu L_\nu$ and segmented into piecewise-constant Bayesian blocks. The peak block is defined as the block with the largest mean $\nu L_\nu$, and $t_{\rm peak}$ is taken to be the midpoint of the earliest and latest measurements contained within this block, indicated by the dashed vertical line.

The TDEs that lie within our selected time window are plotted on the Mollweide skymap in Fig.~\ref{fig:tdes}. We have a total of $N_{\rm TDE} = 89$ TDEs. We classify the TDEs based on their luminosity distance ($d_L$), where the most nearby TDEs are denoted as red circles.
\subsection{HESE neutrino dataset}
\label{subsec:hese}
\begin{figure}
\centering
\includegraphics[width=0.49\textwidth]{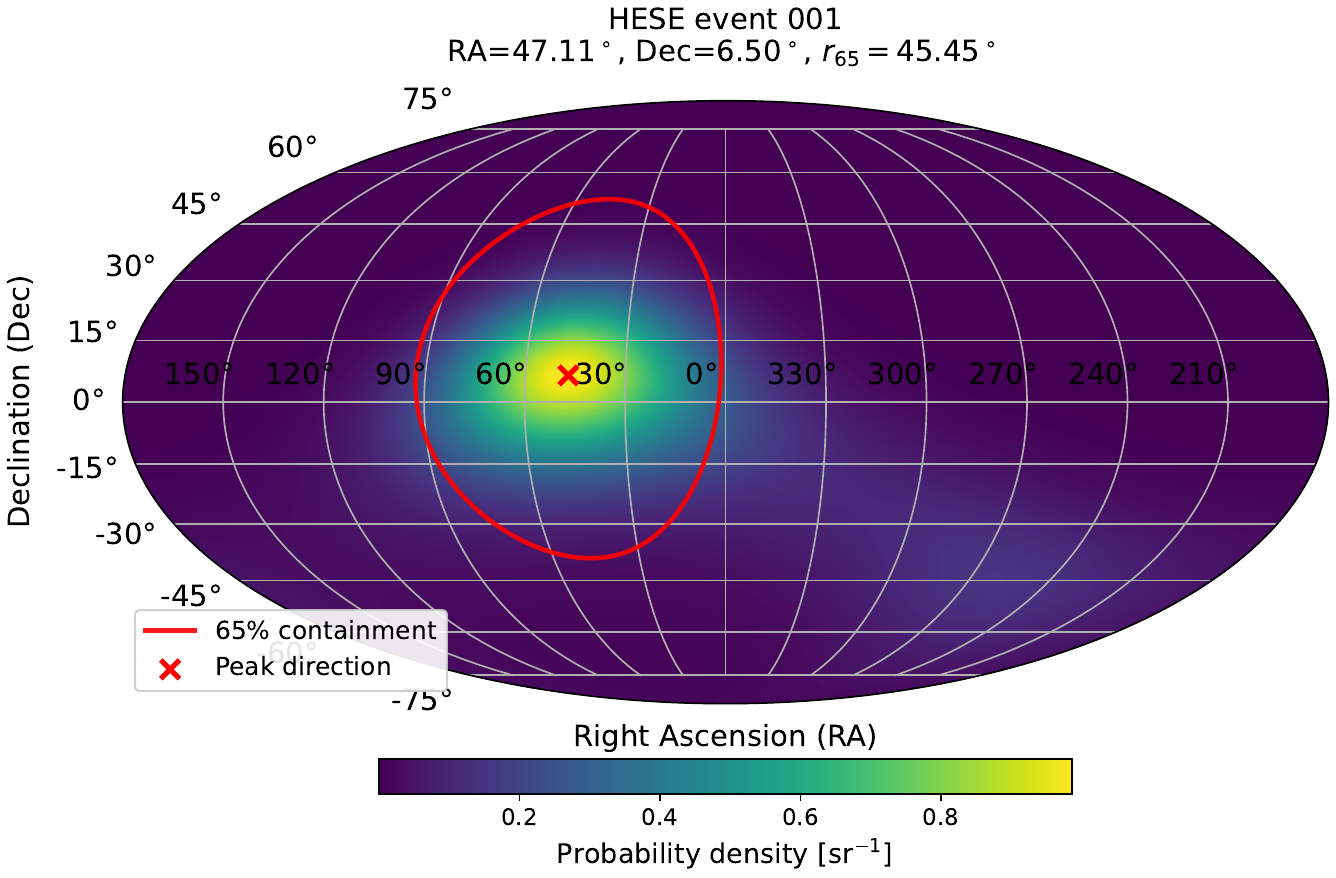}
\includegraphics[width=0.49\textwidth]{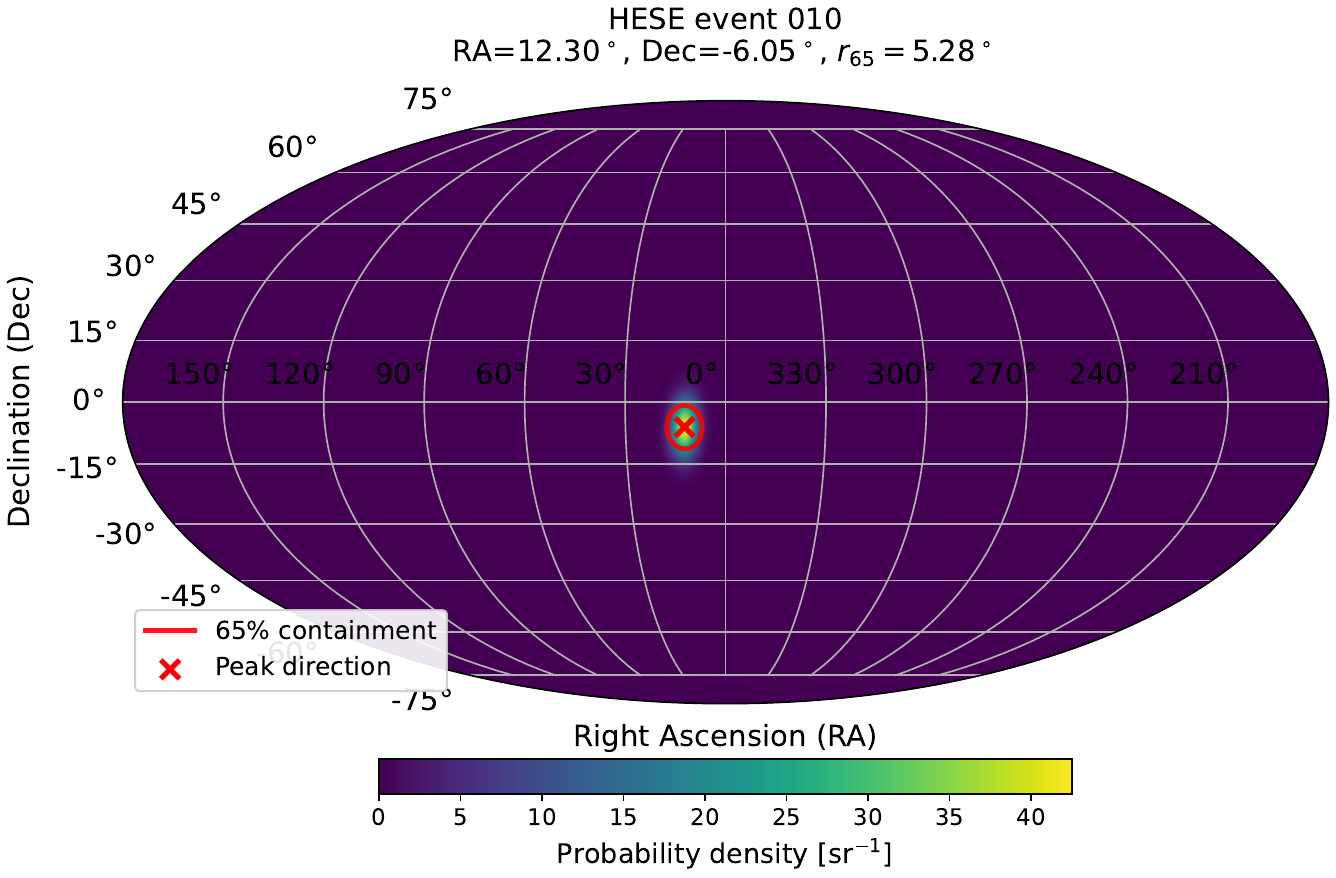}
\caption{\label{fig:hese_error}Mollweide skymaps for two representative HESE events, Event 001 (\emph{left}) and Event 010 (\emph{right}). The color scale represents the directional probability density in each sky pixel. The red cross marks the best-fit event direction, and the red circle denotes the corresponding $65\%$ containment region.
}
\end{figure}
\begin{figure}
\centering
\includegraphics[width=1\linewidth]{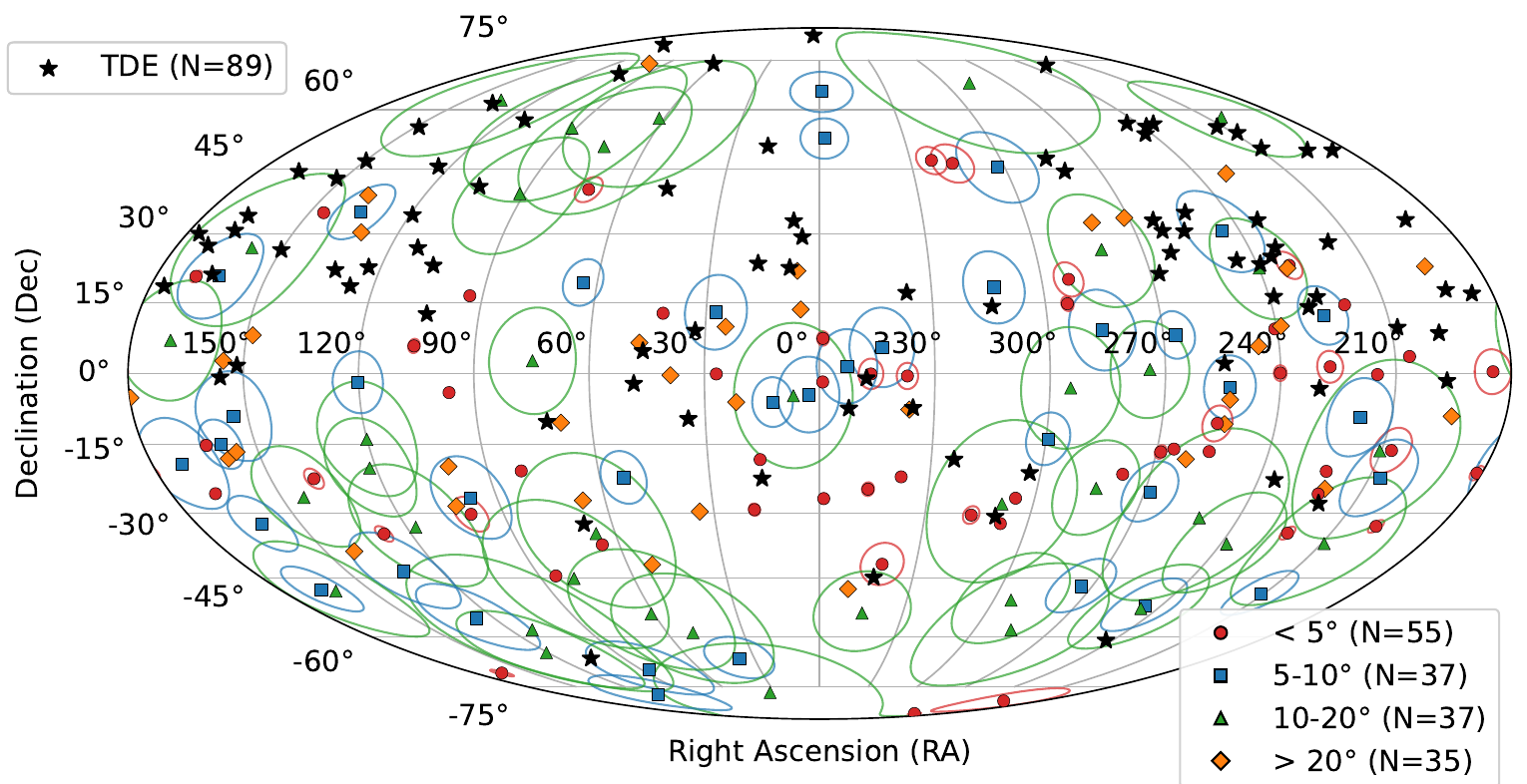}
\caption{\label{fig:hese} Mollweide projection for the HESE~\cite{hesedata, IceCube:2023sov}. We denote the 65\% containment angle for each neutrino event (see the text for details). This is denoted using markers and colors depending on their size.}
\end{figure}
The IceCube Neutrino Observatory is essentially a Cherenkov light detector employing optical sensors containing photomultiplier tubes (PMT), called digital optical modules (DOMs), embedded within the Antarctic ice. Neutrino interactions within the ice or bedrock generate relativistic charged particles that emit Cherenkov light. These interactions fall into two categories: neutral-current (NC) and charged-current (CC). Neutrino NC interactions initiate a hadronic shower that is detected by the DOMs. These interactions are mediated by the $Z$ boson, in which the outgoing neutrino remains unobservable and carries away an unknown amount of energy. In contrast, CC interactions are mediated by the $W$ boson. These interactions also initiate a hadronic shower, however, the neutrino is converted into its corresponding charged lepton. An electromagnetic shower initiates when this outgoing lepton is an electron, which appears cascade-like and makes CC interactions indistinguishable from NC. Alternatively, when the outgoing lepton is a muon, it can traverse several kilometers through the ice. This produces a track-like signature of energy depositions along a roughly fixed direction. Track events have finer angular resolution and hence are appropriate for point source searches, whereas shower events have finer energy resolution.

The IceCube Collaboration characterizes High-Energy Starting Events (HESE)~\cite{IceCube:2020wum} as the highest-energy neutrino interactions initiated within the detector’s fiducial volume, with a lower bound on the total detected charge of $6000$ photoelectrons. HESE samples aim to isolate astrophysical neutrinos by reducing the background of both atmospheric muons and atmospheric neutrinos. In practice, outer layers of the detector are used as a veto, and only events with a contained interaction vertex are selected. The IceCube Collaboration released data from June 2010 to August 2022, which contains 164 reconstructed events designated as HESE. This 12-year data release~\cite{hesedata} consists of 102 events from a previous 7.5-year HESE sample with 62 newly selected events, in which 2 events were excluded as coincident backgrounds (events 128 and 132) \cite{IceCube:2023sov}. Over the course of IceCube's operation, ice modeling techniques have improved with microscopic descriptions of ice anisotropy and layer undulations through ice crystal birefringence. These improvements, along with a greater understanding of detectors, more accurately describe directional reconstruction of events and have been applied to previous HESE samples. 

DirectFit was applied to all 164 events to refine the reconstruction of each event's arrival direction~\cite{IceCube:2023sov}. This was followed by a second Approximate Bayesian Computation to sample the posterior density, along with further marginalization over position and energy of the resulting samples. For each event, this release includes the Modified Julian Date of detection, RA and Dec coordinates of the most-probable arrival direction, and a fitted 8-parameter Fisher-Bingham (FB8) distribution to describe the directional probability density function~\cite{IceCube:2023sov}. This FB8 distribution generalizes the von Mises-Fisher distribution by allowing anisotropic angular structure rather than enforcing circular symmetry~\cite{Yuan:2019dxx}. The additional parameters allow modeling of elongated localization regions while preserving normalization. Using these 8 parameters, we generate a normalized equal-area HEALPix projection, which samples the probability density at the pixel’s center direction and stores this value for the corresponding bin of coordinates. We approximate the probability contained in some pixel by multiplying its density by the pixel’s solid angle.

While generating HEALPix projections for each event, the probabilities contained within each pixel are stored to compute the containment radius. This radius can be used to approximately compute an angular error region around the reconstructed event direction. We do this in the following way. The HESE’s center is set to the RA and Dec coordinates provided in this release. After determining each pixel’s angular distance from the HESE’s center, we sort the pixels by distance and, noting that these are normalized projections, find the first radius where the cumulative sum of probabilities reaches 0.65. This method is merely a crude fit to the containment radius and does not effectively model anisotropic projections, however, these radii are not used in our main analysis. Rather, approximate probabilities are extracted using exact coordinates in this analysis. The resulting skymaps for randomly chosen HESE is shown in Fig.~\ref{fig:hese_error}, where the color map represents the directional probability density in each pixel using which the computed best-fit direction and the $65\%$ containment region is shown using a red cross and a red circle respectively.

For our selected time window, there are a total of $N_\nu = 164$ HESE shown in Fig.~\ref{fig:hese}. Using the algorithm described above, we compute the 65\% error containment angle and divide the events based on their angular error. In particular, we classify the events in angular bins of $< 5^\circ$, $5^\circ - 10^\circ$, $10^\circ - 20^\circ$, and $>20^\circ$. The angular error cones are plotted for all cases where the angular error is less than $20^\circ$. We also show the TDEs in the selected time window with a black star in the same Mollweide skymap.
\section{Unbinned likelihood analysis}
\label{sec:unbinnedL}
In this section we outline the unbinned likelihood analysis method we use for our analysis. In essence, it is similar to the one used in~\cite{Esmaili:2018wnv} but with modifications implemented in both the signal and the background probability distribution functions (PDFs). The likelihood (or log-likelihood) function $\mathcal{L}$ is defined as
\be
\label{eq:logL}
{\rm log}\ \mathcal{L} (n_s) = \sum_{i=1}^{N_{\rm TDE}} {\rm log}\ \left[ \frac{n_s}{N_{\rm TDE}} \mathcal{S}_i + \left( 1 -  \frac{n_s}{N_{\rm TDE}}\right) \mathcal{B}_i \right]\,,
\ee
where $\mathcal{S}_i$ and $\mathcal{B}_i$ are the signal and background PDFs, the total number of TDEs in the catalog is given by $N_{\rm TDE}$, and $n_s$ is the parameter that the likelihood function is extremized over\footnote{Note that in the case of high statistics datasets, $n_s$ mimics the number of signal events.}. Given the form on the likelihood function in the above equation, the corresponding test statistic (TS) is defined in the usual way as
\be
\label{eq:ts}
{\rm TS}(n_s) = 2\ {\rm log} \left(\frac{\mathcal{L}(n_s)}{\mathcal{L}(n_s = 0)}\right)\,.
\ee
The best fit value of the parameter $n_s$ is obtained by extremizing TS. The unbinned likelihood function defined in Eq.~\eqref{eq:logL} stacks all associations of neutrino events in the temporal and spatial vicinity of a given TDE $i$. Therefore the analysis is unbinned as far as the neutrino dataset is concerned. This ansatz works well for low statistic datasets, which is the case for this work. But it is important to note here that Eq.~\eqref{eq:logL} differs from the ansatz where the search for correlation is performed by stacking up all source contributions for a given neutrino event $j$~\cite{IceCube:2017amx,Kheirandish:2019dii,Chang:2022hqj}. Ideally, the two ansatz would agree for highly correlated high statistic datasets.

The signal and the background PDFs are both composed of temporal and spatial parts which contribute independently, such that $\mathcal{S}_i = \mathcal{S}_i^{\rm temp} \times \mathcal{S}_i^{\rm spat}$ and $\mathcal{B}_i = \mathcal{B}_i^{\rm temp} \times \mathcal{B}_i^{\rm spat}$. Ideally, an independent contribution from the energy PDF in both the signal and the background PDFs should be included. However, such an inclusion is not needed for the current analysis given the low statistics.

Assuming that the optical peak of the $i$-th TDE occured at $T_i^{\rm TDE}$ and the arrival time of the $j$-th neutrino event is $T_j^{\nu}$, the temporal PDF ($\mathcal{S}_{ij}^{\rm temp}$) is defined using a Gaussian distribution in $T_j^{\nu} - T_i^{\rm TDE}$
\be
\label{eq:sigtemp}
\mathcal{S}_{ij}^{\rm temp} = \frac{1}{\sqrt{2 \pi \sigma_T^2}} \exp \left[ - \frac{\left((T_j^{\nu} - T_i^{\rm TDE}) - \lambda_T \right)^2}{2 \sigma_T^2} \right]\,,
\ee
where $\lambda_T$ is the mean and $\sigma_T$ gives the spread. Both the parameters are motivated from physical modeling of the TDEs. For example, for the previously claimed neutrino associations, the delay between the optical peak of the TDE and the neutrino arrival was $\mathcal{O}(100)$ days. Irrespective of the validity of such associations, the formation of a radiatively inefficient accretion flow (RIAF) leading to a magnetically arrested accretion disk (MAD) state is possible. This would in turn delay the launching of jets from TDEs~\cite{Tchekhovskoy:2013tt,Mukhopadhyay:2023mld} and, hence, lead to a delayed neutrino arrival since these neutrinos are produced in the external shock of the jet. The spread quantifies the variation of this delay. For a more detailed analysis with more data, precise modeling of $\lambda_T$ and $\sigma_T$ is needed. However, observationally motivated values with plausible physical explanations will suffice for our purposes. Hence, we fix $\lambda_T = 100$ days and $\sigma_T = 10$ days\footnote{See Appendix~\ref{appsec:sigT} for a discussion on alternate choices of $\sigma_T$.}. 

For the background temporal PDF, we assume a uniform distribution within a time window. From Eq.~\eqref{eq:sigtemp}, the $99$\% confidence interval is given by $T_j^{\nu} - T_i^{\rm TDE} \in [\lambda_T-2.58 \sigma_T, \lambda_T + 2.58 \sigma_T]$, which implies the time interval is $2\ 2.58\ \sigma_T \approx 50$ days. Thus, the uniform background PDF is given by
\be
\mathcal{B}_{ij}^{\rm temp} = \frac{1}{2 \times 2.58\ \sigma_T}\,.
\ee
For the $i$-th TDE the contributions from the signal and background temporal PDF, given a neutrino event $j$ are stacked (summed) leading to
\be
\label{eq:si_bi_temp}
\mathcal{S}_i^{\rm temp} = \sum_{j=1}^{N_\nu} \mathcal{S}_{ij}^{\rm temp}\,; \hspace{0.2em} \mathcal{B}_i^{\rm temp} = \sum_{j=1}^{N_\nu} \mathcal{B}_{ij}^{\rm temp}\,,
\ee
where $N_\nu$ is the total number of neutrino events.

\begin{figure}
\centering
\includegraphics[width=0.49\textwidth]{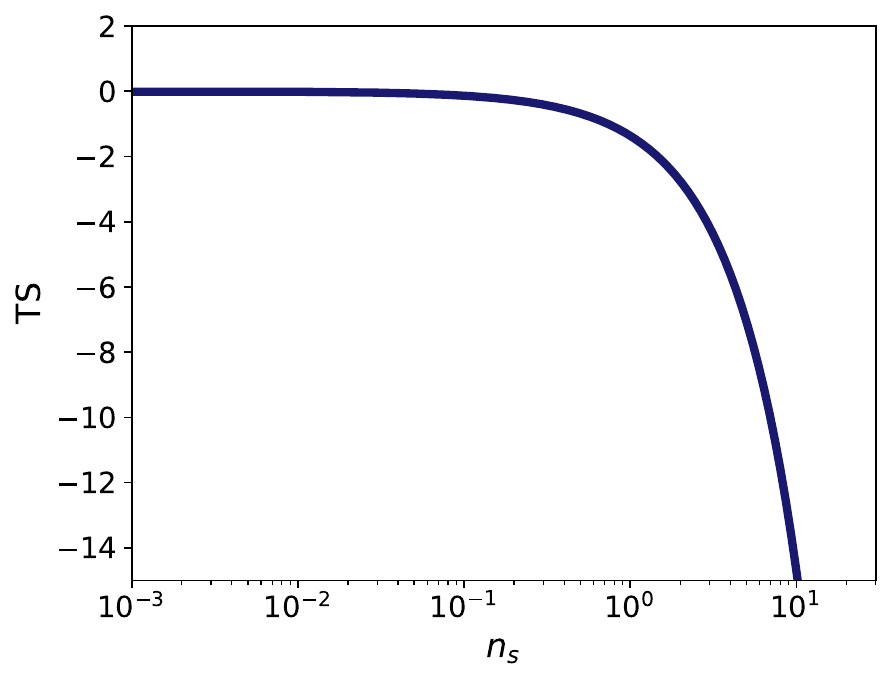}
\includegraphics[width=0.49\textwidth]{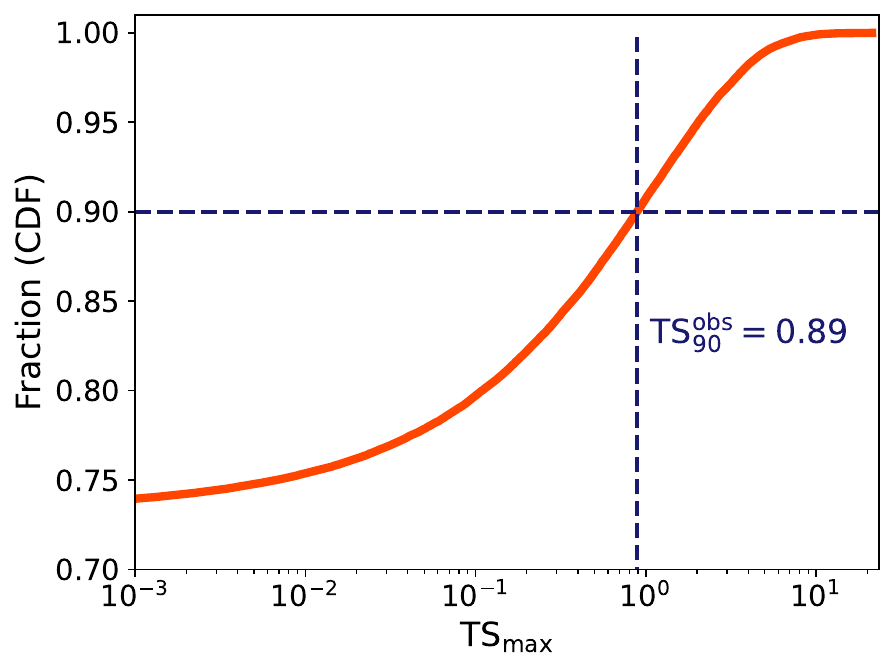}
\caption{\label{fig:ts_cdf} \emph{Left: }Test Statistic TS as a function of signal parameter $n_s$. We find TS$_{\rm max} = 0$ corresponding to $n_s=0$. This implies that the observational data is consistent with background-only hypothesis and shows no correlation. \emph{Right: }The cumulative distribution fraction (CDF) of TS$_{\rm max}$ for $10^5$ sets of randomly generated TDEs. The blue dashed line shows the value of TS$_{90}^{\rm obs} = 0.89$ denoting that $90$\% of TS$_{\rm max}$ values are less than this number.
}
\end{figure}

Next, we evaluate the contributions from the signal and background parts of the spatial PDFs. The current HESE neutrino dataset from IceCube provides the full spatial PDF as discussed in Sec.~\ref{subsec:hese}. This is provided using an 8-parameter fit to the Fisher-Bingham distribution (FB). The FB8 distribution more accurately models the signal PDF for TDEs with asymmetric sky localizations than a circular von Mises-Fisher distribution. For each neutrino event map $j$ given a RA and Dec one can read the value of the PDF from the map. For a given map and its corresponding value of parameter $N_{\rm side}$, where $N_{\rm side}$ is the HEALPix resolution parameter that fixes the total number of equal-area pixels on the sphere as $N_{\rm pix} = 12 N_{\rm side}^2$. For the HESE dataset the skymaps have $N_{\rm side} = 256$. The area per pixel can be defined as $\Omega_{\rm pix} = 4\pi/N_{\rm pix} = \pi/\big( 3 N_{\rm side}^2 \big)$. The spatial signal PDF is then defined in the following way
\be
\mathcal{S}_{ij}^{\rm spat} = \mathcal{A} \Omega_{\rm pix}\,,
\ee
where $\mathcal{A}$ is defined as the probability density per steradian contained in some pixel. The background spatial distribution is assumed to be uniform and is thus given by
\be
\mathcal{B}_{ij}^{\rm spat} = \frac{1}{4\pi} \Omega_{\rm pix}\,,
\ee
Similar to Eq.~\eqref{eq:si_bi_temp} the spatial signal and background PDFs for a given TDE $i$ stacked over all neutrinos is given by, 
\be
\label{eq:si_bi_spat}
\mathcal{S}_i^{\rm spat} = \sum_{j=1}^{N_\nu} \mathcal{S}_{ij}^{\rm spat}\,; \hspace{0.2em} \mathcal{B}_i^{\rm spat} = \sum_{j=1}^{N_\nu} \mathcal{B}_{ij}^{\rm spat}\,.
\ee
We now have all the constituents needed to evaluate Eq.~\eqref{eq:logL} and hence compute the test statistic given the observed data set (TS$_{\rm obs}$) from Eq.~\eqref{eq:ts}. We show our results for the TS in Fig.~\ref{fig:ts_cdf} \emph{(left)} where the variation of TS with the parameter $n_s$ is shown. We do not see any correlation in this case and thus obtain TS$_{\rm max}^{\rm obs} = 0$ for $n_s = 0$. This clearly shows that the observed data is consistent with the background only hypothesis. In Appendix~\ref{appsec:check} we perform some more checks to elaborate on why we do not see a correlation and also perform artificial tests to test the validity of our analysis. We find that most of our $\mathcal{S}_i/\mathcal{B}_i \ll 1$ and hence is peaked towards low values which justifies our findings.

To obtain the associated p-value we do the following. We generate $10^5$ catalogs each containing $N_{\rm TDE}$ number of TDEs. Each TDE has a random RA$\in [0,2\pi]$, cos (Dec) $\in [-1,1]$, time uniformly chosen between $T_{\rm min}^\nu$ and $T_{\rm max}^\nu$ given by the smallest and largest HESE neutrino event timestamps, and luminosity distance $d_L$ randomly sampled from the list of the observed TDE catalog. For each catalog we compute TS$_{\rm max}$. We show the cummulative distribution fraction (CDF) in the right panel of Fig.~\ref{fig:ts_cdf}. The distribution shows that 90\% of the observed TS$_{\rm max}$ values are less than, TS$_{90}^{\rm obs} = 0.89$.
\section{Upper limits on source properties}
\label{sec:lim}
\begin{figure}
\centering
\includegraphics[width=0.75\textwidth]{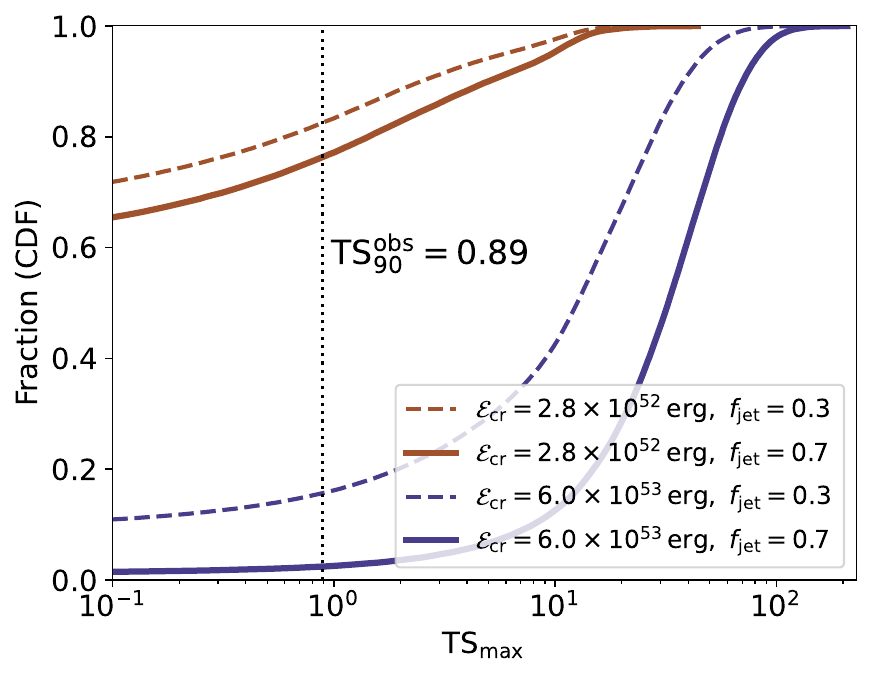}
\caption{\label{fig:ecr_fjet} The CDF of TS$_{\rm max}$ for a combination of $f_{\rm jet}$ and $\mathcal{E}_{\rm CR}$ values. The parameters chosen are $f_{\rm jet} = \{ 0.3, 0.7 \}$ (dashed, solid respectively) and $\mathcal{E}_{\rm CR} = \{ 2.8 \times 10^{52} {\rm erg}, 6.0 \times 10^{53} {\rm erg} \}$ (brown, slate blue respectively). For each parameter set, $10^5$ realizations of signal injected mock TDE catalogs are generated to compute the CDF of TS$_{\rm max}$. The dotted black vertical line shows TS$_{90}^{\rm obs} = 0.89$.
}
\end{figure}
Absence of any correlation from the previous section prevents us from identifying TDEs as sources of neutrinos in the HESE dataset. Nevertheless, the indications from the observational datasets can be used to place constraints on the physical parameters associated with TDEs. As discussed in Sec.~\ref{sec:intro}, the two physical quantities of interest here are $\mathcal{E}_{\rm CR}$ and $f_{\rm jet}$, that is, the total energy deposited in cosmic rays and the fraction of TDEs that are intrinsically jetted respectively. We generate mock TDE catalogs, perform synthetic signal injection, and repeat the steps described above to compute the CDF for the generated TS$_{\rm max}$. The steps to computing this are discussed below.

The fraction of jetted objects, $f_{\rm jet}$, was introduced in Refs~\cite{Senno:2017vtd,Esmaili:2018wnv}. If the emission is isotropic and all sources are neutrino emitters, $f_{\rm jet}=1$ is expected by definition. If all sources have jets, $f_{\rm jet}$ is the same as the beaming factor. In general, only a fraction of the sources may harbor jets intrinsically, in which case $f_{\rm jet}$ is the beaming factor ($f_b$) for intrinsically jetted TDEs. In our case, the apparent rate density of jetted TDEs is $\dot{\rho}^{\rm apparent}_{\rm jetted TDE}\approx f_b\ \dot{\rho}^{\rm true}_{\rm jetted TDE}$ and $\dot{\rho}^{\rm true}_{\rm jetted TDE} \ll \dot{\rho}^{\rm true}_{\rm TDE}$, so we have $f_{\rm jet}=f_b (\dot{\rho}^{\rm true}_{\rm jetted TDE}/\dot{\rho}^{\rm true}_{\rm TDE})$.

We start by generating $N_{\rm TDE}$ number of TDEs. Each TDE has a random RA, Dec, time of optical peak ($T^{\rm TDE}$), and luminosity distance ($d_L$). The RA and Dec are chosen randomly from RA$\in [0,2\pi]$ and cos (Dec) $\in [-1,1]$ respectively, and $T^{\rm TDE}$ is chosen uniformly between minimum and maximum HESE timestamp $T_\nu^{\rm min}$ and $T_\nu^{\rm max}$. Lastly, the luminosity distance is chosen from a scrambled list of luminosity distances from the observed TDE catalog. This is done since the observed sample has a bias to nearby (hence brighter) objects. In our analysis, we do not alter the observed HESE dataset. The synthetic catalogs are generated for the TDE dataset, which is followed by signal injection to the same. This is different from what is typically done, where synthetic neutrino events are simulated to perform correlation searches. Our choice stems from the way we define the likelihood in Sec.~\ref{sec:unbinnedL}.

We fix a value of $f_{\rm jet}$ and $\mathcal{E}_{\rm CR}$ for the remaining steps. The next step involves randomly selecting a subsample of jetted TDEs, $N_{\rm TDE}^{\rm jet} = f_{\rm jet}N_{\rm TDE}$ TDEs. For each of these TDEs (with luminosity distance $d_L$), given a fixed choice of $\mathcal{E}_{\rm CR}$, the total number of expected signal events ($N_\nu^{\rm exp}$) can then be computed using Eq.~\eqref{eq:nusig}, where we set $N_{\rm cat} = N_{\rm TDE}^{\rm jet}$. The total number of signal events ($N_{\rm sig}$) that are to be injected are then sampled from a Poissonian distribution with mean $N_\nu^{\rm exp}$. Next, we select $N_{\rm sig}$ distinct jetted TDEs to correlate with the neutrino events. This is done by implementing weighted selection from the sample of $N^{\rm jet}_{\rm TDE}$ based on the weighting factor $w_i = N_{\nu,i}^{\rm exp}/N_\nu^{\rm exp}$, where $N_{\nu,i}^{\rm exp}$ is the number of neutrino events expected from the $i$-th jetted TDE. This implies that the probability of selecting TDEs that have a higher neutrino yield is higher as one would expect intuitively. We impose two additional conditions which are: (a) each TDE can contribute to at most one signal neutrino association, that is, $N_{\rm sig} \leq N_{\rm TDE}^{\rm jet}$ and (b) each neutrino event can be associated with more than one TDE. The former ensures that we select $N_{\rm sig}$ \emph{distinct} TDEs for signal injection. To enforce this uniqueness the weighting is done from an oversampled\footnote{The oversampling ensures that we do not run out of TDE events to choose from given the requirement of uniqueness.} set of TDE indices, which consists of approximately drawing an oversampled pool of candidate sources with replacement and probabilities proportional to their expected yields ($w_i$), and retaining the first $N_{\rm sig}$ unique indices. This is a valid approximation when $N_{\rm sig} << N_{\rm TDE}^{\rm jet}$, which is the case for us. So at this point we have $N_{\rm sig}$ pairs of TDE and HESE neutrino events.

Let the TDE and HESE neutrino pair be $\left( {\rm TDE}_i,\ {\rm HESE}_j \right)$. The final step consists of artificial signal injection to our dataset. For each of the $N_{\rm sig}$ pairs,  this is done by first choosing the arrival time of HESE$_j$ ($T^{\nu_j}$), then sampling a delay time $\delta T\ (= T^\nu - T^{\rm TDE})$ from the temporal PDF given a $\lambda_T$ and $\sigma_T$ (see Eq.~\ref{eq:sigtemp}). The peak time of the TDE is thus defined as $T^{\rm TDE_i} = T^{\nu_j} - \delta T$. The RA and the Dec are assigned to coincide with the HESE neutrino event. This algorithm ensures perfect spatial correlation, with directional uncertainties accounted for from the HEALPix map. As for the temporal associations, the draw from the temporal signal PDF also ensures highly temporally correlated events.
\begin{figure}
\centering
\includegraphics[width=0.9\textwidth]{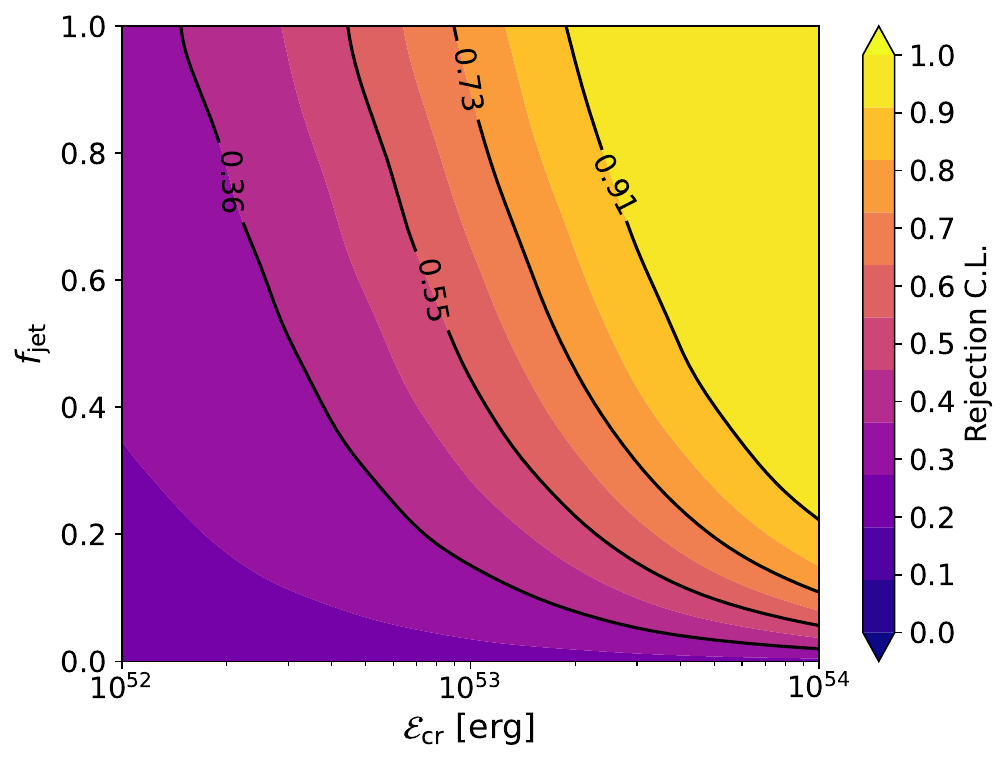}
\caption{\label{fig:uplim} Upper limits in the $f_{\rm jet} - \mathcal{E}_{\rm CR}$ plane for jetted TDEs. The heat map shows the rejection C.L.s along with some contours. Note that even with limited statistics we can constrain the top right corner corresponding to $\mathcal{E}_{\rm CR} \gtrsim 3\times 10^{53}$ erg and $f_{\rm jet} \gtrsim 0.6$ at more than $90$\% C.L. The parameter space of $f_{\rm jet} \gtrsim 0.6$ is relevant for the choked-jet scenarios, while for successful jets $f_{\rm jet} \sim 10^{-4}$.
}
\end{figure}

The signal injected mock TDE catalog is then analyzed using the likelihood analysis discussed in Sec.~\ref{sec:unbinnedL} to compute the associated TS$_{\rm \max}$. The above steps are implemented on $10^5$ synthetically generated and signal injected TDE catalogs for a given value of $f_{\rm jet}$ and $\mathcal{E}_{\rm CR}$ to obtain the CDF of TS$_{\rm max}$. In Fig.~\ref{fig:ecr_fjet} we show the CDF of TS$_{\rm max}$ for a combination of representative values of $f_{\rm jet} = \{ 0.3, 0.7 \}$ (dashed, solid respectively) and $\mathcal{E}_{\rm CR} = \{ 2.8 \times 10^{52} {\rm erg}, 6.0 \times 10^{53} {\rm erg} \}$ (brown, slate blue respectively). The test statistic corresponding to $90$\% fraction of observed TS$_{\rm max}$, TS$_{90}^{\rm obs} = 0.89$ is shown as a dotted black line. 

The rejection confidence level (C.L.) for a chosen value of $f_{\rm jet}$ and $\mathcal{E}_{\rm CR}$ is given by the fraction of TS$_{\rm max}$ values that are more than TS$_{90}^{\rm obs}$. This immediately reflects a few things from Fig.~\ref{fig:ecr_fjet}. The rejection C.L. increases as $f_{\rm jet}$ increases. For example for $\mathcal{E}_{\rm CR} = 2.8 \times 10^{52}\ {\rm erg}$ when $f_{\rm jet}$ changes from $ 0.3$ to $0.7$ the rejection C.L. goes down from $\sim 20$\% to $\sim 30$\%. Similarly for a fixed value of $f_{\rm jet}$ the rejection C.L. decreases in increase in $\mathcal{E}_{\rm CR}$. This is clear from the change of rejection C.L. from $\sim 20$\%  to $\sim 85$\% when 
$\mathcal{E}_{\rm CR}$ changes from $2.8 \times 10^{52}\ {\rm erg}$ to $6.0 \times 10^{53}\ {\rm erg}$ for $f_{\rm jet} = 0.3$.

In Fig.~\ref{fig:uplim} we show the rejection C.L. as a color map on the $f_{\rm jet} - \mathcal{E}_{\rm CR}$ plane. The most prominent feature is towards the top right corner, where we can constrain the region of parameters $\mathcal{E}_{\rm CR} \gtrsim 3 \times 10^{53}$ erg and $f_{\rm jet} \gtrsim 0.6$ at more than $90$\% C.L., even with limited statistics. As expected we are least sensitive in the bottom left region where $f_{\rm jet} \rightarrow 0$ and $\mathcal{E}_{\rm CR} \sim 10^{52}\ {\rm erg}$. We show the contours corresponding to other rejection C.L.s with solid black lines on the plane.

\subsection{Theoretical implications}

The limits we obtain allow us to make some implications on the theoretical modeling of TDEs. The energy budget for cosmic rays and neutrinos  depends on models. In the disk-corona model, a fraction of the accretion energy is transferred to cosmic rays. In the wind and jet model, a fraction of the kinetic or magnetic energy is expected to be carried by cosmic rays. In either case, we expect that the available energy may be limited by the gravitational binding energy of the bound debris, which is
\be
\label{eq:ebind}
{\mathcal E}_{g} \approx \frac{1}{2} \frac{G M_{\rm BH} M_*}{R_T} \sim 4 \times 10^{53}\ {\rm erg} \left( \frac{M_{\rm BH}}{10^8 M_\odot} \right)^{2/3} \left( \frac{M_*}{1 M_\odot} \right)^{4/3} \left( \frac{R_*}{1 R_\odot} \right)^{2/3}\,.
\ee
Now assuming a fraction $\epsilon_{\rm CR}$ of ${\mathcal E}_{g}$ goes to cosmic rays, we have, ${\mathcal E}_{\rm CR}^j = \epsilon_{\rm CR} {\mathcal E}_{g}$, where the total energy of cosmic rays ${\mathcal E}_{\rm CR}^{j} = f_{b} \mathcal{E}_{\rm CR}$ (recall that $\mathcal{E}_{\rm CR}$ is the isotropic equivalent cosmic-ray energy). Thus one can constrain $\epsilon_{\rm CR}$ using
\be
\label{eq:epsilon_cr}
\epsilon_{\rm CR} \sim \left(\frac{f_{b}}{{\mathcal E}_{g}}\right) \left( \mathcal{E}_{\rm CR} \right)\,,
\ee
From observations we have $f_{\rm jet} \sim 0.01-0.1\%$ for successful (jet-loud) TDE jets. This is because, $f_{\rm jet}$ can also be written as $f_{\rm jet} \sim f_b \times P(\rm jet) \sim 10^{-3} \times 10^{-2} \sim 10^{-4}$, where for typical jet-half opening angle $\theta_j \sim 5^\circ$, $f_b \sim 10^{-3}$, and the observed fraction of jetted TDEs $P(\rm jet) \sim 1\%$. Testing such values ($f_{\rm jet} \sim 10^{-4}$) is out of reach for the currently available data sets as can be seen from the lower left corner of Fig.~\ref{fig:uplim}. This is to be expected since the TDE catalog we use does not have a jet-loud TDE contained in it. However with upcoming data from the Vera C. Rubin Observatory, lower values of $f_{\rm jet}$ can also be probed using the methods outlined in our work. On the other hand, for $f_{\rm jet} = 1$, that is, if all TDEs are neutrino emitters and the emission is isotropic, we can constrain $\mathcal{E}_{\rm CR} < 2 \times 10^{53}$ erg at 90\% C.L. (upper right corner of Fig.~\ref{fig:uplim}). This also by definition implies $f_{\rm jet} = f_b = 1$ and hence we can constrain $\epsilon_{\rm CR} \lesssim 0.5$ at 90\% C.L. assuming a 1 $M_\odot$ star and $10^8 M_\odot$ SMBH.

However, it is important to note that the above two possibilities for $f_{\rm jet}$ serve as the limiting boundaries. The value of $f_{\rm jet}$ can be much larger than the observational value of $f_{\rm jet} \sim 0.01-0.1\%$ for successful (jet-loud) TDE jets. This is possible if TDEs intrinsically harbor jets which are launched with a delay and therefore are more prone to get choked~\cite{Mukhopadhyay:2023mld}. In such cases, $f_{\rm jet}\gg 0.1\%$ and our results from Fig.~\ref{fig:uplim} are useful. For choked jets the emission is isotropic and there is no suppression from the beaming factor. For example, in the case of tidal disruption of a 1 $M_\odot$ star and $10^8 M_\odot$ SMBH, the allowed region at 90\% from Fig.~\ref{fig:uplim}, $\mathcal{E}_{\rm CR} < 3 \times 10^{53}$ erg and $f_{\rm jet} < 0.6$, implies we have conservative \emph{constraints} with $\epsilon_{\rm CR} \lesssim 0.75 f_b$, where $f_{\rm jet}<f_b$ (implying large beaming angle) should be satisfied. This allows for scenarios that are non-jetted and have isotropic neutrino emission, that is $f_{b} \sim 1$. In particular, the hidden wind, choked jet~\cite{Mukhopadhyay:2023mld,Lu:2023miv}, and the corona model are all viable for high energy neutrino emission from TDEs (see~\cite{Murase:2020lnu} for details).

The upper limits on the all-sky diffuse neutrino flux ($E_\nu^2\Phi_\nu$) can also be computed from our results in Fig.~\ref{fig:uplim}. This can be estimated analytically using~\cite{Murase:2015xka}
\begin{eqnarray}
E_\nu^2&\Phi_\nu& \approx \frac{c t_H}{4\pi} \varepsilon_\nu Q_{\varepsilon_\nu} \xi_z 
\approx \frac{c t_H}{4\pi} \left(\frac{3}{8}\right) \xi_z \frac{f_{\rm jet} \dot{\rho}^{\rm true}_{\rm TDE} \mathcal{E}_{\rm CR}}{\mathcal{R}}\,,\\
&\lesssim&\;
1.6\times10^{-7}\,
\mathrm{GeV}~\mathrm{cm^{-2}\,s^{-1}\,sr^{-1}}\,
\left(\frac{\xi_z}{0.5}\right)
\left(\frac{f_{\rm jet}}{1.0}\right)
\left(\frac{\dot{\rho}^{\rm true}_{\rm TDE}}{100\,\mathrm{Gpc^{-3}\,yr^{-1}}}\right)
\left(\frac{\mathcal{E}_{\rm CR}}{2\times10^{53}\,\mathrm{erg}}\right)
\left(\frac{18}{\mathcal{R}}\right)\,\nonumber
\end{eqnarray}
where the limits are at the 90\% C.L. In the above equation, $c$ is the speed of light, $t_H$ is the age of universe ($=4.35 \times 10^{17}$ s), the factor of $4\pi$ accounts for the solid angle. The factor $3/8$ comes from considering high energy neutrino production through $p\gamma$ processes which is dominant in TDE environments. The redshift evolution factor for TDE luminosity density is given by $\xi_z$, $\dot{\rho}^{\rm true}_{\rm TDE}$ is the true rate density for TDEs, and $\mathcal{R}$ is the conversion factor between bolometric and differential luminosities. The limits for the all-sky diffuse neutrino flux are indeed weak, that is, it is much larger than the observed all-sky diffuse all flavor flux by IceCube ($\sim 10^{-8}\ \mathrm{GeV}\mathrm{cm^{-2}\,s^{-1}\,sr^{-1}}$). However, with more data from various current and upcoming surveys (such as the Vera C. Rubin Observatory), smaller values of $f_{\rm jet}$ can be effectively constrained, making the limits competitive.

Besides the stacking limits, complementary constraints can also be derived from the non-detection of multiplet sources (bright and nearby sources which produce more than one detected neutrino events from the same sky direction within observation time $T_{\rm IC}$)~\cite{Murase:2016gly,Senno:2016bso}
\be
\dot{\rho}_0^{\rm eff} \gtrsim 1.4 \times 10^{4}\ {\rm Gpc}^{-3}{\rm yr}^{-1} 
\left(\frac{b_m q_L}{6.6}\right)^{2}
\left(\frac{T_{\rm IC}}{6\ {\rm yrs}}\right)^{2}
\left(\frac{\Delta \Omega}{2 \pi}\right)^{2}
\left(\frac{\xi_z}{0.5}\right)^{-3}
\left(\frac{\phi_{\rm lim}}{10^{-0.9}}\right)^{-3}\,,
\ee
where $\dot{\rho}_0^{\rm eff}$ is the effective local ($z=0$) rate density, $b_m q_L$ is a correction factor that depends on the details of the analysis (see~\cite{Murase:2016gly} for details), $T_{\rm IC}$ is the observation time which is the number of years for which the observed IceCube data is being used, $\Delta \Omega$ is the field of view of the detector, and $\phi_{\rm lim}$ is the neutrino fluence limit. We see that $\dot{\rho}_0^{\rm eff} >> \dot{\rho}_{\rm TDE}^{\rm on-axis}$ which implies that jetted (radio-loud) TDEs can only contribute by a subdominant fraction to the all-sky diffuse flux. However, choked jets leading to a large value of $f_{\rm jet}$ are still allowed even when placing constraints in this way.
\section{Summary and outlook}
\label{sec:concl}
TDEs can provide sites of efficient particle acceleration and hence serve as high energy neutrino sources. Coincident neutrino events from TDEs have been searched for and in some cases probable hints of the same were found~\cite{Stein:2020xhk,Reusch:2021ztx,vanVelzen:2021zsm,Yuan:2024foi}. Based on observational evidence, roughly $1$\% of the TDE population are thought to harbor jets~\cite{Burrows:2011dn, Zauderer:2011mf,DeColle:2019wzp,Cenko:2011ys,Brown:2015amy}. Both successful and choked jets in TDEs can produce neutrinos. Furthermore, late time radio observations indicate a rise in the radio light curves $100 - 1000$ days post the optical peak, hinting towards late time central engine activity~\cite{Horesh:2021gvp,Cendes:2023rrc,Sato:2024zpq,Wu:2025vsr}.

The IceCube Collaboration provides a sample of high energy neutrinos where the event is contained in the detector volume, that is, the interaction vertex is reconstructed to be inside the detector volume, which also ensures excellent atmospheric background suppression. This sample, called the HESE sample, offers an opportunity to look for associations with transients including supernovae and TDEs. In this work, we use the 12.5 year HESE sample~\cite{hesedata} along with a TDE catalog provided in~\cite{Langis:2025btl} to search for temporal and spatial correlations using the unbinned maximum likelihood method (see Sec.~\ref{sec:unbinnedL}). Our sample consists of $89$ TDEs and $164$ HESE within the overlapping time window (see Figs.~\ref{fig:tdes} and~\ref{fig:hese}). We perform a version of stacking search where for a given TDE we stack its association in the temporal and spatial domain with all HESE neutrino events. Our analysis shows that the current datasets are consistent with the background-only hypothesis and we do not find any significant correlation between them (Fig.~\ref{fig:ts_cdf}).

The absence of significant correlation allowed us to place upper limits on two physical quantities of interest the isotropic-equivalent energy in cosmic rays ($\mathcal{E}_{\rm CR}$) and the fraction of TDEs that intrinsically harbor jets ($f_{\rm jet}$). We generate synthetic datasets and perform artificial signal injection (see Sec.~\ref{sec:lim}) which allows us to show constraints in the $f_{\rm jet} - \mathcal{E}_{\rm CR}$ plane (Fig.~\ref{fig:uplim}) using the 90-th percentile of test statistic (TS) from the observed dataset. Most interestingly, even with limited observational statistics, the constraints we obtain are $\mathcal{E}_{\rm CR} \lesssim 3 \times 10^{53}$ erg for $f_{\rm jet} \gtrsim 0.6$ at more than $90$\% C.L. We discuss the theoretical implications of our results focusing on the energy budget in cosmic rays and evaluating the diffuse all-sky neutrino flux limits. Furthermore, we comment on the viability of various TDE neutrino production models and sites. Our results clearly demonstrate that using the same techniques but with more statistics in the future, stringent constraints can be placed on physical quantities associated with TDEs, which is important for gaining insights into understanding TDEs from a combination of observational data.

Needless to say with more observational data, our present analysis will need to be refined at various stages. In the present work, we do not assume an energy PDF in the likelihood function. The inclusion of a physically motivated neutrino energy spectrum template instead of assuming a generic $E_\nu^{-2}$ template, will further help with extending the correlation search to the energy domain besides the existing spatial and temporal ones. For placing the upper limits too, instead of the currently used $E_\nu^{-2}$ template a theory informed template can provide more meaningful constraints. Furthermore, the temporal signal PDF can also be improved by choosing $\lambda_T$ and $\sigma_T$ from observational data (see Appendix~\ref{appsec:sigT}). Our current choice of $\lambda_T \approx 100$ days was motivated by prior claims of TDE-neutrino associations, which has become uncertain~\cite{IceCube:2025uzh}. Moreover, there exists a wide diversity of TDEs and hence the choice of $\sigma_T = 10$ days is on the narrower side. On one hand, such a choice of a narrower temporal PDF strengthens the search results by reducing the accumulated background, on the other hand it also introduces a bias by suppressing possible associations with substantially earlier or later neutrino arrival times. A broader temporal PDF, or an event-dependent choice of $\sigma_T$ inferred from individual TDE light curves (as discussed in Appendix~\ref{appsec:sigT}), would generally increase the background contribution and weaken the sensitivity. Furthermore, in our Bayesian Blocks approach (described in Sec.~\ref{subsec:tde_cat}), the $t_{\rm peak}$ estimates can be biased, often towards later times, for events with incomplete temporal coverage around the peak. We leave such an extended temporal treatment, including profiling or marginalization over $\lambda_T$ and $\sigma_T$, for future work with larger TDE and neutrino samples.

It is also important to note that in the current analysis we did not mask out the IceCube neutrino events originating from the direction of the Galactic Plane~\cite{IceCube:2023ame}. The impact of this effect on the present search is expected to be limited for two reasons. First, the Galactic-plane emission is a diffuse component that is approximately steady on the timescales relevant for this analysis, whereas the TDE signal hypothesis requires both spatial coincidence and temporal coincidence with the assumed delayed emission window after the optical peak. Second, the observed dataset gives TS$_{\max}^{\rm obs}=0$, indicating that no positive spatial-temporal correlation is found. Thus, the Galactic-plane contribution does not appear to produce a spurious TDE-neutrino excess in the current dataset. A more complete treatment in future work, especially with larger TDE and neutrino samples, should include a direction-dependent background PDF or dedicated Galactic-plane masking tests.

Improved directional reconstruction techniques from IceCube~\cite{IceCube:2025uzh,IceCube:2023sov} along with better angular resolution from KM3NeT~\cite{KM3Net:2016zxf} and Baikal-GVD (Gigaton Volume Detector)~\cite{Baikal-GVD:2018isr} will improve the results for the spatial PDF. This will complement the amount of neutrino data available from upcoming neutrino telescopes like GRAND~\cite{GRAND:2018iaj,Kotera:2025jca}, RNO-G~\cite{RNO-G:2020rmc}, Pacific Ocean Neutrino Experiment (P-ONE)~\cite{P-ONE:2020ljt}, TRIDENT~\cite{TRIDENT:2022hql}, Trinity~\cite{Otte:2025dld}, and Tau Air-shower Mountain-Based Observatory (TAMBO)~\cite{TAMBO:2025jio}. 
Besides the plethora of data available from the current EM telescopes, the upcoming ones like The Square Kilometre Array Observatory (SKAO)~\cite{Donnarumma:2015bra} in the radio band, LSST~\cite{LSST:2008ijt} and Ultraviolet Transient Astronomy Satellite (ULTRASAT)~\cite{Shvartzvald:2023ofi} in the optical/UV band, X-Ray Imaging and Spectroscopy Mission (XRISM)~\cite{XRISMScienceTeam:2020rvx} and Einstein Probe (EP)~\cite{Yuan:2025cbh} in X-rays, All-sky Medium Energy Gamma-ray Observatory eXplorer (AMEGO-X)~\cite{Caputo:2022xpx}, and Compton Spectrometer and Imager (COSI)~\cite{Tomsick:2023aue} in MeV gamma rays, and Fermi Large
Area Telescope (LAT)~\cite{Peng:2016nfz} and Cherenkov Telescope Array Observatory (CTAO)~\cite{CTAConsortium:2017dvg} for high energy gamma rays will greatly add to the catalogs of TDEs available in each band to search for correlations. The sheer statistics from observations in the near future will greatly enhance the scope of the current analysis presented in this work.
\acknowledgments
We would like to thank Arman Esmaili, Nicholas Kamp, and Karri I. I. Koljonen for helpful discussions and invaluable suggestions. M.\,M. wishes to thank the organizers of WHEPP 2025 where the final stages of this work were completed.
M.\,M. acknowledges support from the FermiForward Discovery Group, LLC under Contract No. 89243024CSC000002 with the U.S. Department of Energy, Office of Science, Office of High Energy Physics and the Institute for Gravitation and the Cosmos (IGC) Postdoctoral Fellowship. The work of K.M. is supported by the NSF Grants, No.~AST- 2108466, No.~AST-2108467 and No.~AST-2308021.
\appendix
\section{Choice of $\sigma_T$ using the TDE catalog}
\label{appsec:sigT}
\begin{figure}
\centering
\includegraphics[width=0.75\linewidth]{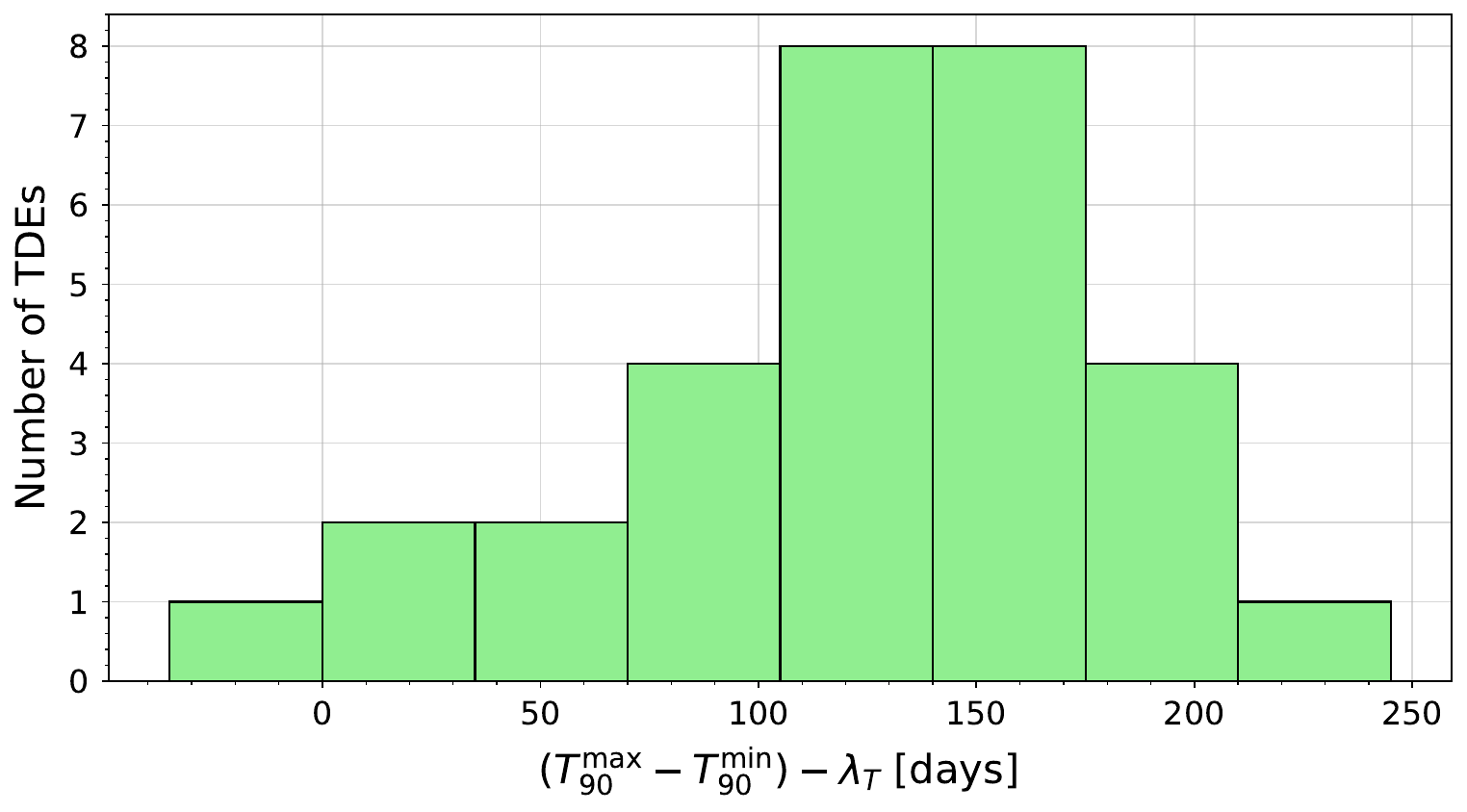}
\caption{\label{fig:sigt_dist} Histogram showing the $\left(T_{90}^{\rm max} - T_{90}^{\rm min}\right) - \lambda_T$ distribution for a subset of $30$ TDEs~\cite{Hammerstein:2022wia} from our analysis.}
\end{figure}
The choice of $\sigma_T$ can be improved by considering the light curves associated with the TDE events. A prescription to do this is as follows. Given the light curve in the optical band, we can compute the time stamps between which $90$\% of the area is covered. Assume we assign them values $T_{90}^{\rm min}$ and $T_{90}^{\rm max}$ (technically we enforce a lower and upper bound on the integration limits such that $-50\ {\rm days} < T_{90}^{\rm min} < T_{90}^{\rm max} < 350\ {\rm days}$), such that the former is prior to the optical peak and the latter is post the optical peak. One can then compute $\left(T_{90}^{\rm max} - T_{90}^{\rm min}\right) - \lambda_T$ for each TDE in the catalog. This quantity can then be plotted as a histogram to obtain a mean value for the spread, which would serve as $\sigma_T$. Choosing a larger value of $\sigma_T$ leads to an increase in the background accumulated and, hence, reduces $\mathcal{S}_i/\mathcal{B}_i$.

The histogram for a subset of our TDE catalog is shown in Fig.~\ref{fig:sigt_dist}. We select a sample of $30$ TDEs that lie within our selected time window. Choosing a bin of $35$ days we approximately find a Gaussian distribution for $\sigma_T$ with a peak between $100 - 135$ days. Thus, with a larger TDE dataset, the choice of $\lambda_T$ and $\sigma_T$ can also be statistically sampled from the observed data.
\section{Validity of analysis}
\label{appsec:check}
\begin{figure}
\centering
\includegraphics[width=0.49\textwidth]{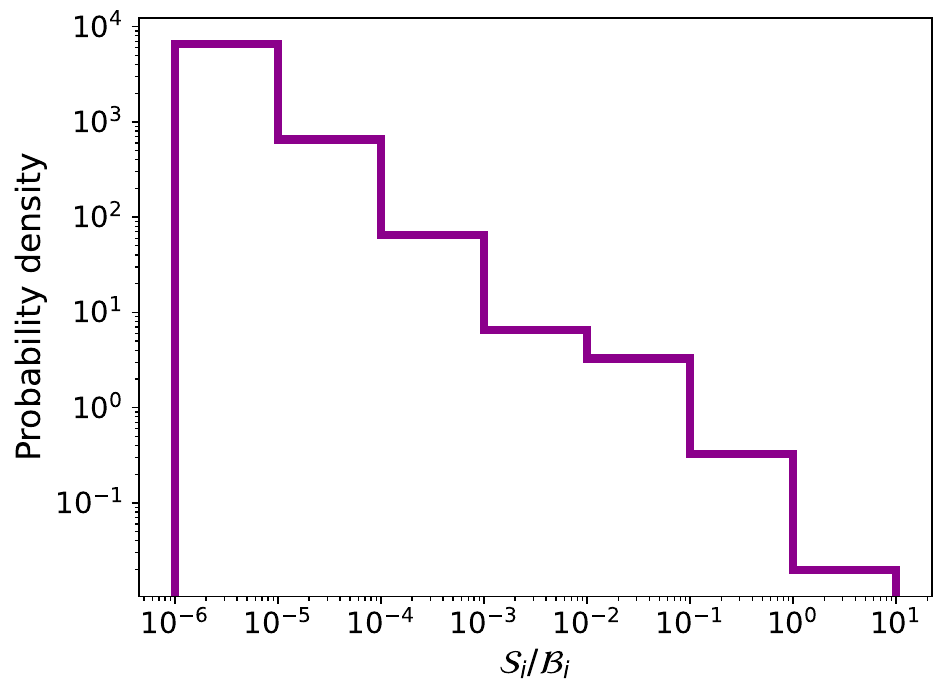}
\includegraphics[width=0.49\textwidth]{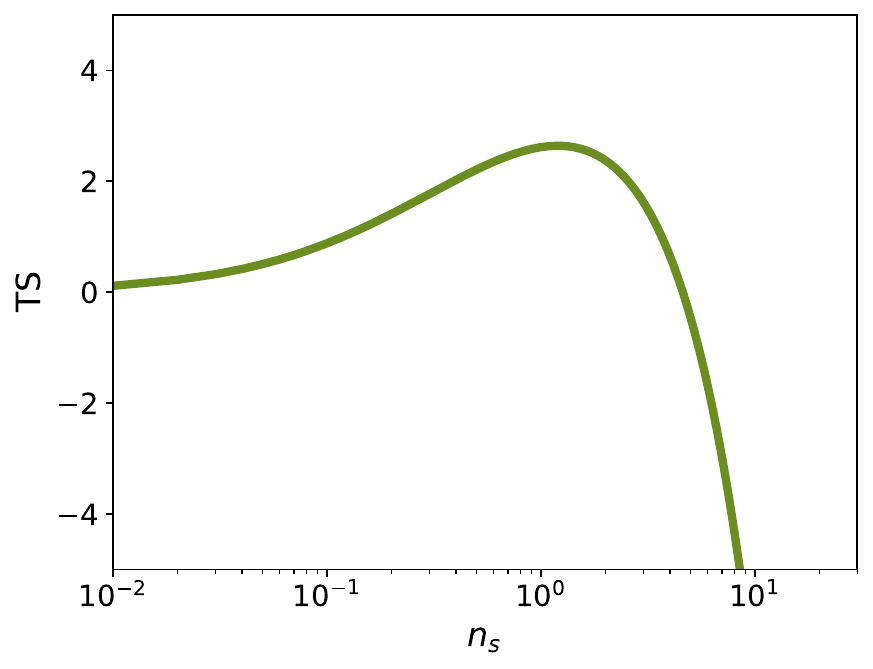}
\caption{\label{fig:tests}\emph{Left:} The probability density distribution for $\mathcal{S}_i/\mathcal{B}_i$ from the observed datasets, where $\mathcal{S}_i$ and $\mathcal{B}_i$ are computed using the procedure outlined in Sec.~\ref{sec:unbinnedL}. The pile up in the probability density at lower values of $\mathcal{S}_i/\mathcal{B}_i$ is consistent with the bakcground only hypothesis and the corresponding values of TS$_{\rm max}$ and $n_s$ at TS$_{\rm max}$ we obtain (see Fig.~\ref{fig:ts_cdf}). \emph{Right:} Test Statistic TS as a function of signal parameter $n_s$ when one of the $\mathcal{S}_i/\mathcal{B}_i$ is artificially boosted by 100 times its original value. We find TS$_{\rm max} = 2.64$ corresponding to $n_s=1.2$.
}
\end{figure}
Since our results for the observed datasets are consistent with the background only hypothesis, it is imperative we perform some consistency and sanity checks to test the validity of our analysis. In this Appendix we present some results justifying our analysis.

In Fig.~\ref{fig:tests} \emph{(left)} we plot the probability density $p(w) dw$ where $w_i = \mathcal{S}_i/\mathcal{B}_i$, that is, the fraction of events with $w \in [w,w + dw]$ and normalized such that $\int p(w) dw = 1$. We note that there is a big pile up in the probability density for $\mathcal{S}_i/\mathcal{B}_i \sim 10^{-6} - 10^{-5}$. This is consistent with our conclusion that most of the correlations we have are background like (of course with signal injection and introduction of artificial correlations this changes as evident from Sec.~\ref{sec:lim}). In fact most of our TS with the observed dataset is driven by the tail of this distribution, where, $\mathcal{S}_i/\mathcal{B}_i \sim 1 - 10$. Since for most events in our dataset $\mathcal{S}_i/\mathcal{B}_i << 1$, these events disfavor $n_s>0$ in the likelihood, causing it to decrease as we move away from $n_s \rightarrow 0$. Thus what we obtain as our result in Sec.~\ref{sec:unbinnedL} and Fig.~\ref{fig:ts_cdf} can be explained by this. In fact, this is why without artificial signal injection with just the background only Monte Carlo realizations, we still find a very low value of  TS$_{\rm max}^{\rm obs} = 0.89$.

To further test the validity of our results, we can artificially boost $\mathcal{S}_i/\mathcal{B}_i$ for some event $i$. We choose the event with the highest  $\mathcal{S}_i/\mathcal{B}_i$ and boost it by a factor of 100, while keeping the rest of the signal and background PDFs the same. On evaluating the likelihood and TS on this modified list of $\mathcal{S}_i$ and $\mathcal{B}_i$, we find that the distribution of TS with $n_s$ shown in Fig.~\ref{fig:tests} \emph{(right)} gives TS$_{\rm max} = 2.64$ corresponding to $n_s=1.2$. This is still consistent with the background only hypothesis but highlights that our result of TS$_{\rm max} = 0$ and $n_s = 0$ is indeed an effect of $\mathcal{S}_i/\mathcal{B}_i << 1$ and if indeed our dataset had signal like events, that is, $\mathcal{S}_i/\mathcal{B}_i >> 1$ we would favor larger values of $n_s>0$.
\section{Photometric band selection for TDE peak-time approximation}
\label{appsec:tpk_tde_catalog}
In this subsection, we provide details about the TDEs used in our analysis. The full sample of 134 TDEs can be found in Ref.~\cite{Langis:2025btl}. We summarize the information about the telescope used, the band selected, and the $t_{\rm peak}$ value obtained by implementing the Bayesian Blocks segmentation in Table~\ref{tab:tde_band_grouping}.

We first select a single photometric band within a window spanning 50 days before and 300 days after the discovery date. Preferably, we chose the Zwicky Transient Facility's (ZTF)~\cite{Bellm_2019} $g$-band, requiring a minimum of 5 measurements within the time window. The ZTF uses 48-inch Samuel Oschin Telescope at Palomar Observatory to repeatedly image large areas of the sky in optical $g$, $r$, and $i$ bands, producing transient light curves through reference-image difference imaging and point spread function (PSF)-based photometry. If the above requirement is not satisfied, we chose the ZTF $r$-band. If neither ZTF bands satisfy this requirement, we select the filter with the most observations. The Asteroid Terrestrial-impact Last Alert System's (ATLAS) is an optical survey with intensive temporal sampling using optical $c$ and $o$ bands~\cite{Tonry}. This facility measures transient brightness primarily through image-subtraction photometry. The All-Sky Automated Survey for SuperNovae's (ASASSN) is a network of small telescopes using optical $g$ and Johnson-Cousins $V$ bands~\cite{Holoien, Christy}. This network provides transient light curves through calibrated photometric reductions~\cite{Kochanek_2017}. The Catalina Real-Time Transient Survey's (CRTS) is an optical survey implementing Catalina survey imaging using standard photometric calibration practices to measure transient fluxes~\cite{Drake}. The Palomar Transient Factory (PTF) provides optical transient surveys, mainly in the $R$ band, with the Palomar 48-inch telescope to measure transients using automated image subtraction and PSF photometry~\cite{Law, Masci}.

Other bands used include the Las Cumbres Observatory (LCO) $B$ band, the Catalina Sky Survey's (CSS) $V$ band, the Panoramic Survey Telescope and Rapid Response System 1's (PS1) $g$ band. LCO is a global network of robotic telescopes providing optical imaging in standard filters with photometrically calibrated reductions~\cite{Brown}. CSS is an optical survey optimized for moving near-Earth objects that provides time-series photometry in a V-like band through repeated imaging and calibrated photometric reductions~\cite{Drake}. PS1 is another optical survey that provides transient light curves through repeatedly imaging, primarily with its $g_{P1}$ filter, and calibrated multi-epoch photometry~\cite{Chambers, Schlafly}. There are 3 events (AT2022cmc, Swift J2058+08, and Swift J1644+57) that do not satisfy this requirement for any single filter and are excluded from this analysis.

\begingroup
\scriptsize
\setlength{\parindent}{0pt}
\setlength{\columnsep}{0.9cm}

\newcommand{\TDErow}[3]{%
\noindent
\makebox[0.27\linewidth][l]{#1}%
\makebox[0.35\linewidth][l]{#2}%
\makebox[0.30\linewidth][r]{#3}%
\par
}

\newcommand{\TDErule}{%
\vspace{1pt}
\noindent\rule{0.94\linewidth}{0.35pt}
\vspace{1pt}
\par
}

\captionsetup{type=table}
\caption{
Table of the photometric band used for each TDE peak-time approximation provided by the Bayesian-block analysis. The source catalog is provided in Ref.~\cite{Langis:2025btl}.
}
\label{tab:tde_band_grouping}

\begin{paracol}{2}

\TDErule
\TDErow{Band}{TDE name}{$t_{\rm peak}$ [MJD]}
\TDErule

\TDErow{$g$--ZTF}{AT2018bsi}{58229.7186}
\TDErow{$g$--ZTF}{AT2018hco}{58433.7820}
\TDErow{$g$--ZTF}{AT2018iih}{58461.8217}
\TDErow{$g$--ZTF}{AT2018jbv}{58482.0187}
\TDErow{$g$--ZTF}{AT2018lna}{58503.2159}
\TDErow{$g$--ZTF}{AT2018lni}{58474.1143}
\TDErow{$g$--ZTF}{AT2018mli}{58508.4929}
\TDErow{$g$--ZTF}{AT2018zr}{58204.2145}
\TDErow{$g$--ZTF}{AT2019azh}{58561.7316}
\TDErow{$g$--ZTF}{AT2019bhf}{58547.9397}
\TDErow{$g$--ZTF}{AT2019cho}{58550.3069}
\TDErow{$g$--ZTF}{AT2019dsg}{58617.9651}
\TDErow{$g$--ZTF}{AT2019ehz}{58619.2891}
\TDErow{$g$--ZTF}{AT2019gte}{58644.2010}
\TDErow{$g$--ZTF}{AT2019lwu}{58694.4057}
\TDErow{$g$--ZTF}{AT2019meg}{58696.8095}
\TDErow{$g$--ZTF}{AT2019mha}{58698.6947}
\TDErow{$g$--ZTF}{AT2019qiz}{58763.4586}
\TDErow{$g$--ZTF}{AT2019teq}{58830.0972}
\TDErow{$g$--ZTF}{AT2019vcb}{58828.5511}
\TDErow{$g$--ZTF}{AT2020acka}{59221.0596}
\TDErow{$g$--ZTF}{AT2020afhd}{59180.2126}
\TDErow{$g$--ZTF}{AT2020ddv}{58909.2348}
\TDErow{$g$--ZTF}{AT2020ksf}{59024.3467}
\TDErow{$g$--ZTF}{AT2020mbq}{59025.7130}
\TDErow{$g$--ZTF}{AT2020mot}{59066.9364}
\TDErow{$g$--ZTF}{AT2020neh}{59031.1981}
\TDErow{$g$--ZTF}{AT2020nov}{59093.7071}
\TDErow{$g$--ZTF}{AT2020ocn}{59033.2444}
\TDErow{$g$--ZTF}{AT2020opy}{59061.7295}
\TDErow{$g$--ZTF}{AT2020pj}{58862.9777}
\TDErow{$g$--ZTF}{AT2020qhs}{59095.3395}
\TDErow{$g$--ZTF}{AT2020riz}{59083.9602}
\TDErow{$g$--ZTF}{AT2020ukj}{59215.1457}
\TDErow{$g$--ZTF}{AT2020vdq}{59131.9718}
\TDErow{$g$--ZTF}{AT2020vwl}{59209.5560}
\TDErow{$g$--ZTF}{AT2020wey}{59152.0141}
\TDErow{$g$--ZTF}{AT2020ysg}{59151.5309}
\TDErow{$g$--ZTF}{AT2020yue}{59182.9824}
\TDErow{$g$--ZTF}{AT2020zso}{59190.6522}
\TDErow{$g$--ZTF}{AT2021ack}{59239.4953}
\TDErow{$g$--ZTF}{AT2021axu}{59251.4548}
\TDErow{$g$--ZTF}{AT2021crk}{59274.2130}
\TDErow{$g$--ZTF}{AT2021ehb}{59296.1833}
\TDErow{$g$--ZTF}{AT2021gje}{59318.8927}
\TDErow{$g$--ZTF}{AT2021jjm}{59331.3356}
\TDErow{$g$--ZTF}{AT2021jsg}{59272.1637}
\TDErow{$g$--ZTF}{AT2021lo}{59261.6967}
\TDErow{$g$--ZTF}{AT2021mhg}{59383.9337}
\TDErow{$g$--ZTF}{AT2021nwa}{59405.7510}
\TDErow{$g$--ZTF}{AT2021qth}{59401.8455}
\TDErow{$g$--ZTF}{AT2021qxv}{59370.7848}
\TDErow{$g$--ZTF}{AT2021sdu}{59413.4369}
\TDErow{$g$--ZTF}{AT2021uqv}{59450.4182}
\TDErow{$g$--ZTF}{AT2021utq}{59460.1887}
\TDErow{$g$--ZTF}{AT2021uvz}{59491.1424}
\TDErow{$g$--ZTF}{AT2021yte}{59477.4768}
\TDErow{$g$--ZTF}{AT2021yzv}{59510.9337}
\TDErow{$g$--ZTF}{AT2022aee}{59625.7094}
\TDErow{$g$--ZTF}{AT2022arb}{59631.3376}
\TDErow{$g$--ZTF}{AT2022bdw}{59628.3478}
\TDErow{$g$--ZTF}{AT2022csn}{59654.1728}
\TDErow{$g$--ZTF}{AT2022dbl}{59639.3561}
\TDErow{$g$--ZTF}{AT2022dsb}{59689.4605}
\TDErow{$g$--ZTF}{AT2022dyt}{59676.3114}
\TDErow{$g$--ZTF}{AT2022emf}{59625.8260}
\TDErow{$g$--ZTF}{AT2022exr}{59674.4457}

\TDErule

\switchcolumn

\TDErule
\TDErow{Band}{TDE name}{$t_{\rm peak}$ [MJD]}
\TDErule

\TDErow{$g$--ZTF}{AT2022fpx}{59803.2695}
\TDErow{$g$--ZTF}{AT2022gdw}{59692.8248}
\TDErow{$g$--ZTF}{AT2022gri}{59967.3518}
\TDErow{$g$--ZTF}{AT2022hvp}{59697.7634}
\TDErow{$g$--ZTF}{AT2022ibq}{59694.9483}
\TDErow{$g$--ZTF}{AT2022lri}{59899.2312}
\TDErow{$g$--ZTF}{AT2022pna}{59794.4205}
\TDErow{$g$--ZTF}{AT2022rz}{59600.4311}
\TDErow{$g$--ZTF}{AT2022upj}{59900.1943}
\TDErow{$g$--ZTF}{AT2022wtn}{59908.1239}
\TDErow{$g$--ZTF}{AT2023clx}{60287.4651}
\TDErow{$g$--ZTF}{AT2023cvb}{60016.4525}
\TDErow{$g$--ZTF}{AT2023kvy}{60112.4515}
\TDErow{$g$--ZTF}{AT2023mfm}{60152.3843}
\TDErow{$g$--ZTF}{AT2023mhs}{60135.2186}
\TDErow{$g$--ZTF}{AT2023rvb}{60213.1670}
\TDErow{$g$--ZTF}{AT2023tmq}{60244.0049}
\TDErow{$g$--ZTF}{AT2023vto}{60241.1884}
\TDErow{$g$--ZTF}{AT2023wdb}{60247.1926}
\TDErow{$g$--ZTF}{AT2023xen}{60260.4061}
\TDErow{$g$--ZTF}{AT2024admt}{60670.4315}
\TDErow{$g$--ZTF}{AT2024aepd}{60680.5355}
\TDErow{$g$--ZTF}{AT2024agqx}{60706.9411}
\TDErow{$g$--ZTF}{AT2024as}{60327.8319}
\TDErow{$g$--ZTF}{AT2024bgz}{60360.7591}
\TDErow{$g$--ZTF}{AT2024eqp}{60406.7242}
\TDErow{$g$--ZTF}{AT2024gre}{60449.2336}
\TDErow{$g$--ZTF}{AT2024gxr}{60462.3535}
\TDErow{$g$--ZTF}{AT2024kt}{60339.2579}
\TDErow{$g$--ZTF}{AT2024lhc}{60532.7502}
\TDErow{$g$--ZTF}{AT2024mvz}{60482.4380}
\TDErow{$g$--ZTF}{AT2024pvu}{60543.3273}
\TDErow{$g$--ZTF}{AT2024qab}{60525.7264}
\TDErow{$g$--ZTF}{AT2024rny}{60550.8577}
\TDErow{$g$--ZTF}{AT2024tvd}{60579.1896}
\TDErow{$g$--ZTF}{AT2024ule}{60574.4993}
\TDErow{$g$--ZTF}{AT2024vip}{60608.4092}
\TDErow{$g$--ZTF}{AT2024wsd}{60600.3976}
\TDErow{$g$--ZTF}{AT2024ymu}{60628.6983}
\TDErow{$g$--ZTF}{AT2024yqo}{60630.4539}
\TDErow{$g$--ZTF}{AT2024zau}{60639.8512}
\TDErow{$r$--ZTF}{AT2022adm}{59600.4941}
\TDErow{$r$--ZTF}{AT2023lli}{60406.0116}
\TDErow{$r$--ZTF}{AT2023ugy}{60393.5248}
\TDErow{$o$--ATLAS}{AT2017eqx}{57927.0589}
\TDErow{$o$--ATLAS}{AT2017gge}{57997.2528}
\TDErow{$o$--ATLAS}{AT2018hyz}{58456.6121}
\TDErow{$o$--ATLAS}{AT2022agi}{59559.2812}
\TDErow{$o$--ATLAS}{AT2022czy}{59650.3345}
\TDErow{$o$--ATLAS}{AT2024doo}{60342.3263}
\TDErow{$c$--ATLAS}{iPTF16axa}{57519.0015}
\TDErow{$g$--ASAS-SN}{AT2018dyb}{58343.0711}
\TDErow{$g$--ASAS-SN}{AT2018fyk}{58427.6002}
\TDErow{$g$--ASAS-SN}{AT2019ahk}{58547.0602}
\TDErow{$g$--ASAS-SN}{AT2021blz}{59364.6970}
\TDErow{$V$--ASAS-SN}{AT2016fnl}{57635.4147}
\TDErow{CL--CRTS}{ASASSN-14li}{57103.3053}
\TDErow{CL--CRTS}{PTF09djl}{55092.1506}
\TDErow{CL--CRTS}{iPTF15af}{57161.9590}
\TDErow{$R$--PTF}{ASASSN-14ae}{56316.9257}
\TDErow{$R$--PTF}{PTF09ge}{54994.8223}
\TDErow{$B$--ASAS-SN}{ASASSN-15oi}{57250.7660}
\TDErow{$V$--CSS}{LSQ12dyw}{56199.7372}
\TDErow{$g$--PS1}{PS1-10jh}{55414.8100}
\TDErow{--}{AT2022cmc}{--}
\TDErow{--}{Swift J2058+08}{--}
\TDErow{--}{Swift J1644+57}{--}

\TDErule

\end{paracol}
\endgroup
\bibliography{refs}
\bibliographystyle{jhep}
\end{document}